\documentclass[12pt]{article}
\pdfoutput=1
\usepackage[T1]{fontenc}
\usepackage{graphicx,psfrag,epsf,color}
\usepackage{amsmath,amssymb,amsfonts}
\usepackage{physics}
\usepackage{siunitx}
\usepackage{array}
\usepackage{cite}
\usepackage{rotating}
\usepackage{slashed, cancel}
\usepackage{xcolor}
\usepackage{hyperref}
\usepackage[caption=false]{subfig}
\usepackage{graphicx}
\usepackage{mathrsfs}
\usepackage{float}
\restylefloat{table}

\usepackage{tabularx}
\usepackage{booktabs}
\usepackage{manfnt} 
\usepackage{subfig}
\usepackage[margin=1.0in]{geometry}
\newcolumntype{C}{>{\centering\arraybackslash}X}
\usepackage{multirow} \usepackage{array}
\title{}
\author{andersrehult}
\date{}

\begin{document}


\begin{titlepage}

\vspace*{-2.0truecm}

\begin{flushright}
Nikhef-2022-022\\
SI-HEP-2022-36\\
P3H-22-123
\end{flushright}

\vspace*{1.3truecm}

\begin{center}
{
\Large \bf \boldmath Fingerprinting CP-violating New Physics with $B \to K\mu^+\mu^-$}
\end{center}
\vspace{0.9truecm}

\begin{center}
{\bf Robert Fleischer\,${}^{a,b}$,  Eleftheria Malami\,${}^{a,c}$, Anders Rehult\,${}^{a}$, and  K. Keri Vos\,${}^{a,d}$}

\vspace{0.5truecm}

${}^a${\sl Nikhef, Science Park 105, NL-1098 XG Amsterdam,  Netherlands}

${}^b${\sl  Department of Physics and Astronomy, Vrije Universiteit Amsterdam,\\
NL-1081 HV Amsterdam, Netherlands}

${}^c${\sl  Center for Particle Physics Siegen (CPPS), Theoretische Physik 1,\\
Universität Siegen, D-57068 Siegen, Germany}

{\sl $^d$Gravitational 
Waves and Fundamental Physics (GWFP),\\ 
Maastricht University, Duboisdomein 30,\\ 
NL-6229 GT Maastricht, the
Netherlands}\\[0.3cm]

\end{center}

\vspace*{1.7cm}


\begin{abstract}
\noindent
The $B \to K \ell^+ \ell^-$ system is particularly interesting to explore physics beyond the Standard Model. Such effects could provide 
new sources of CP violation, thereby giving rise to Wilson coefficients which could not only be real, as assumed in most analyses, but also be complex. We propose a new strategy to extract the complex $C_{9 \mu}$, $C_{10 \mu}$ from the data, utilising the complementarity of direct CP violation in $B^+\to K^+ \mu^+\mu^-$ and mixing-induced CP violation arising -- in addition to direct CP violation -- in $B^0_d \to K_{S} \mu^+\mu^-$, as well as the differential rate in appropriate $q^2$ bins. The long-distance effects, which play a key role to generate the direct CP asymmetries, cannot be reliably calculated due to their non-perturbative nature. In order to describe them, we apply a model by the LHCb Collaboration employing experimental data. Assuming specific scenarios, we demonstrate the fingerprinting of CP-violating New Physics and the distinction between ambiguities in the model of the long-distance contributions. Finally, New Physics could couple differently to muons and electrons, as probed through the lepton flavour universality ratio $R_K$. We discuss these effects in the presence of CP violation and present a new way to measure the direct CP asymmetries of the $B \to K e^+ e^-$ channels.

\end{abstract}


\vspace*{2.1truecm}

\vfill

\noindent
December 2022

\end{titlepage}

\thispagestyle{empty}

\vbox{}



\setcounter{page}{0}

\newpage
\section{Introduction}
The charged and neutral $B \to K \mu^+ \mu^-$ decays are powerful probes to test the Standard Model of particle physics (SM) and to search for New Physics (NP). For several years, these decays and similar processes originating from the same $b\to s\mu^+\mu^-$ quark-level transition have exhibited interesting tensions between the SM predictions and experimental data \cite{LHCb:2014cxe,LHCb:2014vgu,LHCb:2021trn,BELLE:2019xld,BaBar:2012mrf}. 

A multitude of global effective field theory (EFT) analyses have been performed to determine the short-distance physics that could cause these anomalies (see e.g.~\cite{Bobeth:2017vxj,Altmannshofer:2021qrr,Alguero:2022wkd,Gubernari:2022hxn,Geng:2021nhg,Carvunis:2021jga,Mahmoudi:2022hzx,SinghChundawat:2022zdf}). The EFT framework allows for model-independent analyses in terms of operators and their short-distance Wilson coefficients which -- in most analyses -- are considered to be real. However, in general, NP could also provide new sources of CP violation, which are encoded in complex phases of Wilson coefficients. 
While the usual key players in the search for new CP violation are non-leptonic $B$ decays \cite{Fleischer:2002ys,Fleischer:2022axm}, CP violation might also enter leptonic \cite{Fleischer:2017ltw} and semileptonic rare $B$ decays \cite{Bobeth:2011gi,Alok:2011gv,Becirevic:2020ssj,Bordone:2021olx,Descotes-Genon:2020tnz,Descotes-Genon:2015hea}. 

For the charged $B^\pm \to K^\pm \mu^+\mu^-$ modes, direct CP violation could arise from the interference of two amplitudes with different CP-conserving and CP-violating phases. In the decays at hand, the CP-conserving phases arise through $\bar{c}c$ resonances. In the SM, the CP-violating phases come from elements of the unitary Cabibbo--Kobayashi--Maskawa (CKM) matrix.
For the decays we study here, the matrix element $|V_{cb}|$ is particularly relevant. Unfortunately, we are facing discrepancies between inclusive and exclusive determinations of this quantity, which has implications for SM calculations of the decay rate. In our analysis, we pay special attention to these determinations and address them separately, in line with \cite{DeBruyn:2022zhw}. Neglecting tiny doubly Cabibbo-suppressed terms, the direct CP asymmetry is zero. Consequently, a nonzero value would be an unambiguous sign of new sources of CP violation.

A qualitative NP analysis of direct CP violation thus crucially relies on the description of resonance effects \cite{Becirevic:2020ssj}.
These non-perturbative, i.e.\ long-distance effects are challenging to compute (see e.g.\cite{Khodjamirian:2012rm, Lyon:2013gba, Ciuchini:2015qxb, Capdevila:2017ert, Gubernari:2022hxn}). 
Here we use a long-distance model implemented by the LHCb collaboration based on the Kr\"uger--Seghal approach \cite{Kruger:1996cv, Kruger:1996dt}. Performing a fit to their experimental data, four sets of fit parameters (branches) were obtained \cite{LHCb:2016due}.

In order to obtain the full picture, we also consider the neutral $B_d^0\to K_S\mu^+\mu^-$ decay. In this channel, in addition to direct CP violation, there is also mixing-induced CP violation arising from interference between $\bar{B}^0_d$--$B^0_d$ mixing and $B_d^0, \bar{B}_d^0$ decays into the $K_S\mu^+\mu^-$ final state. In contrast to direct CP violation, mixing-induced CP violation does not require CP-conserving phases. Therefore it is much more robust with respect to long-distance effects \cite{Descotes-Genon:2020tnz}. 

In this paper, we present a new strategy that exploits the complementary information provided by the direct and mixing-induced CP asymmetries to determine the complex phases of the Wilson coefficients $C_{9\mu}$ and $C_{10\mu}$. We demonstrate that different sources of new CP-violating physics leave distinct ``fingerprints'' on the observable space, allowing us to transparently determine the Wilson coefficients. This is possible even without making specific assumptions, such as having NP only in $C_{9\mu}$ or assuming the relation $C_{9\mu}^{\rm NP}=-C_{10\mu}^{\rm NP}$ as is frequently done in the literature. In order for the strategy to work, we need to discriminate between the four long-distance branches. We show how we can do this by considering the CP asymmetries in specific $q^2$ bins.

A first constraint on the direct CP asymmetry has been obtained by the LHCb Collaboration \cite{LHCb:2014mit}. However, the mixing-induced CP asymmetry in $B_d^0\to K_S\ell^+\ell^-$ has not yet been measured. Consequently, we consider benchmark scenarios to illustrate the procedure.

CP-violating NP may also have different strengths for muons versus electrons, which can be probed through the lepton-flavour universality ratio $R_K$ \cite{Bobeth:2007dw,Hiller:2014yaa,Robinson:2021cws}. When allowing for CP-violating effects, special care has to be taken by measuring these ratios for the $B^- (\bar{B}^0_d)$ and $B^+ ({B}^0_d)$ modes separately. In this paper, we define several new observables and find a new way to measure the direct CP asymmetry in $B\to K e^+ e^-$.

This paper is organized as follows: In Section \ref{ch:B_to_Kellell_theoretical_framework}, we introduce the theoretical framework and observables, as well as the model for the long-distance effects. We calculate the SM prediction for the branching ratio of $B^\pm \to K^\pm \mu^+\mu^-$ and, while doing so, address the uncertainty arising from discrepancies between different determinations of the CKM element $|V_{cb}|$. Then, in Section \ref{ch:correlations}, we discuss how we can fingerprint NP with direct and mixing-induced CP asymmetries in $B\to K \mu^+\mu^-$. We discuss the interplay of these observables and demonstrate how we can use them to extract the complex values of $C_{9\mu}$ and $C_{10\mu}$. 
In Section \ref{ch:RK}, we discuss lepton flavour universality. Finally, we conclude in Section \ref{ch:conclusions} .

\section{Theoretical framework}
\label{ch:B_to_Kellell_theoretical_framework}
\subsection{Effective Hamiltonian}
The low-energy effective Hamiltonian for $b \to s \ell^+\ell^-$ transitions is \cite{Descotes-Genon:2020tnz, Altmannshofer:2008dz, Gratrex:2015hna,Buchalla:1995vs} 
\begin{equation}\label{eq:ham}
    \mathcal{H}_{\rm eff} = - \frac{4 G_F}{\sqrt{2}} \left[\lambda_u \Big\{C_1 (\mathcal{O}_1^u - \mathcal{O}_1^c) + C_2 (\mathcal{O}_2^u - \mathcal{O}_2^c)\Big\} + \lambda_t \sum\limits_{i \in I} C_i \mathcal{O}_i \right] \ ,
\end{equation}
where $\lambda_q = V_{qb} V_{qs}^*$ and $I = \{1c, 2c, 3, 4, 5, 6, 8, 7^{(\prime)}, 9^{(\prime)}\ell, 10^{(\prime)}\ell, S^{(\prime)}\ell, P^{(\prime)}\ell, T^{(\prime)}\ell\}$. We neglect the terms proportional to $\lambda_u$ which are doubly Cabibbo suppressed and contribute at the $\mathcal{O}(\lambda^2) \sim 5\%$ level, with $\lambda\equiv |V_{us}|$ in the Wolfenstein expansion of the CKM matrix \cite{Wolfenstein:1983yz,Buras:1994ec}. 
We consider the following operators\footnote{In comparison with \cite{Descotes-Genon:2020tnz}, we have absorbed a factor $m_b$ into the definition of $\mathcal{O}_{S\ell}$ and $\mathcal{O}_{P\ell}$. }:
\begin{equation}
    \begin{aligned}
    \mathcal{O}_{7^{(\prime)}} &= \frac{e}{(4\pi)^2} m_b [\bar s \sigma^{\mu\nu} P_{R(L)} b] F_{\mu\nu}, & \mathcal{O}_{S^{(\prime)}\ell} &= \frac{e^2}{(4\pi)^2} m_b [\bar s P_{R(L)} b] (\bar \ell \ell),\\
    \mathcal{O}_{9^{(\prime)}\ell} &= \frac{e^2}{(4\pi)^2} [\bar s \gamma^\mu P_{L(R)} b] (\bar \ell \gamma_\mu \ell), & \mathcal{O}_{P^{(\prime)}\ell} &= \frac{e^2}{(4\pi)^2} m_b [\bar s P_{R(L)} b] (\bar \ell \gamma_5 \ell),\\
    \mathcal{O}_{10^{(\prime)}\ell} &= \frac{e^2}{(4\pi)^2} [\bar s \gamma^\mu P_{L(R)} b] (\bar \ell \gamma_\mu \gamma_5 \ell), & \mathcal{O}_{T^{(\prime)}\ell} &= \frac{e^2}{(4\pi)^2} [\bar s \sigma^{\mu\nu} P_{R(L)} b] (\bar \ell \sigma_{\mu\nu} \ell) \ ,
    \end{aligned}
\end{equation}
with $P_{R(L)} = \frac{1}{2} (1 \pm \gamma_5)$ and $\sigma_{\mu\nu} = \frac{i}{2}[\gamma_\mu, \gamma_\nu]$. The index $\ell$ denotes the flavour of the lepton, that we will omit for simplicity whenever it is clear from the context if we consider a generic lepton or a specific flavor. In this paper, we will consider only light leptons, i.e.\ $\ell = \mu, e$.  

We use the following standard parameterization for the local form factors,
\begin{align}\label{eq:localff}
    \mel{K^-(p_K)}{\bar s \gamma_\mu b}{B^- (p_B)} &= (p_B + p_K)_\mu f_+(q^2) + \frac{m_B^2 - m_K^2}{q^2} q_\mu (f_0(q^2) - f_+(q^2)), \nonumber\\
     \mel{K^-(p_K)}{\bar s \sigma_{\mu\nu} b}{B^- (p_B)} &= i\bigg{[} (p_B + p_K)_\mu q_\nu - (p_B - p_K)_\nu q_\mu \bigg{]} \frac{f_T(q^2)}{m_B + m_K}, \nonumber \\
      \mel{K^-(p_K)}{\bar s b}{B^- (p_B)} &= \frac{m_B^2 - m_K^2}{m_b - m_s} f_0(q^2),
\end{align}
where $q = p_B-p_K$ is the momentum transfer of the final state lepton-antilepton pair.  

The angular distribution for $B^- \to K^- \ell^+\ell^-$ takes the form (see e.g. \cite{Descotes-Genon:2020tnz, Gratrex:2015hna,Bobeth:2012vn})
\begin{equation}
\begin{aligned}
    \frac{d \Gamma(B^- \to K^-\ell^+\ell^-)}{dq^2 d\cos\theta_\ell} &= & \bar{G}_0(q^2) + \bar{G}_1(q^2) \cos\theta_\ell + \bar{G}_2(q^2)\frac{1}{2} \left(3\cos\theta_\ell^2 -1\right) \ ,
\end{aligned}
\label{eq:BtoKellell_diff_decay_ratecos}
\end{equation}
where $\theta_\ell$ is defined as the angle between the $\ell^-$ three-momentum and the reversed of the $B^-$ three-momentum in the dilepton rest frame.  Integrating \eqref{eq:BtoKellell_diff_decay_ratecos} over the full range $\theta_\ell \in [0,\pi]$, gives 
\begin{equation}
\begin{aligned}
    \frac{d \Gamma(B^- \to K^-\ell^+\ell^-)}{dq^2} &= &2\bar{G}_0(q^2) \ 
\end{aligned}
\label{eq:BtoKellell_diff_decay_rate}
\end{equation}
with
\begin{equation}
\begin{aligned}
    \bar{G}_0 &= & \frac{4}{3}(1 + 2 \hat m_\ell^2) \abs{\bar h_V}^2 + \frac{4}{3} \beta_\ell^2 \abs{\bar h_A}^2 + 2 \beta_\ell^2 \abs{\bar h_S}^2 + 2 \abs{\bar h_P}^2\\
    &&+ \frac{8}{3}(1 + 8 \hat m_\ell^2) \abs{\bar h_{T_t}}^2 + \frac{4}{3}\beta_\ell^2 \abs{\bar h_T}^2 + 16 \hat m_\ell \Im[\bar h_V \bar h_{T_t}^*] \ ,
\end{aligned}
\label{eq:G0bar}
\end{equation}
where $\hat m_\ell \equiv m_\ell/\sqrt{q^2}$ and $\beta_\ell \equiv \sqrt{1 - 4 \hat m_\ell^2}$ \cite{Descotes-Genon:2020tnz}. The coefficients $\bar{G}_1$ and $\bar{G}_2$ are given in \cite{Descotes-Genon:2020tnz}. The decay rate is a function of the amplitudes
\begin{align}
    \bar h_V &= \mathcal{N} \frac{\sqrt{\lambda_B}}{2 \sqrt{q^2}} \bigg[ \frac{2 m_b}{m_B + m_K} (C_7 + C_{7'}) f_T(q^2) + ((C_9 + C_{9'})) f_+(q^2) \bigg], \label{eq:h_amplitudes_V}\\
    \bar h_A &= \mathcal{N} \frac{\sqrt{\lambda_B}}{2 \sqrt{q^2}} (C_{10}+C_{10'}) f_+(q^2),\\
    \bar h_S &= \mathcal{N} \frac{m_B^2 - m_K^2}{2} \frac{m_b}{m_b - m_s} (C_S + C_{S'}) f_0(q^2),\\
    \bar h_P &= \mathcal{N} \frac{m_B^2 - m_K^2}{2} \bigg[ \frac{m_b (C_P + C_{P'})}{m_b - m_s} + \frac{2 m_\ell}{q^2} (C_{10} + C_{10'}) \bigg] f_0(q^2),\\
    \bar h_T &= -i \mathcal{N} \frac{\sqrt{\lambda_B}}{\sqrt{2} (m_B + m_K)} (C_T - C_{T'}) f_T(q^2),\\
    \bar h_{T_t} &= -i \mathcal{N} \frac{\sqrt{\lambda_B}}{2 (m_B + m_K)} (C_T + C_{T'}) f_T(q^2),
    \label{eq:h_amplitudes_Tt}
\end{align}
where $\lambda_B = \lambda(m_B^2, m_K^2, q^2)$ is the Källén function and the normalization factor is
\begin{equation}\label{eq:curlyN}
    \mathcal{N} = - \frac{\alpha G_F}{\pi} V_{ts}^* V_{tb} \sqrt{\frac{q^2 \beta_\ell \sqrt{\lambda_B}}{2^{10} \pi^3 m_B^3}} \ .
\end{equation}

The differential $B^+ \to K^+ \ell^+\ell^-$ decay rate follows from~\eqref{eq:BtoKellell_diff_decay_rate} by replacing $\bar h$ with $h$. The latter are obtained from $\bar h$ by taking the complex conjugate of all CP-violating phases, present in the CKM matrix elements in $\mathcal{N}$ and (possibly) in NP contributions to the Wilson coefficients.

In the remainder of this paper, we neglect possible NP contributions to chirality-flipped (primed) Wilson coefficients such that $h_T= \sqrt{2} h_{T_t}$. Primed Wilson coefficients can be easily included by replacing $C_i \to C_i + C_{i^\prime}$, except in the case of $C_{T}$. 

\subsection{Hadronic long-distance effects}
\label{ch:long_distance_effects}
\begin{figure}
    \centering
    \includegraphics[width=0.4\textwidth]{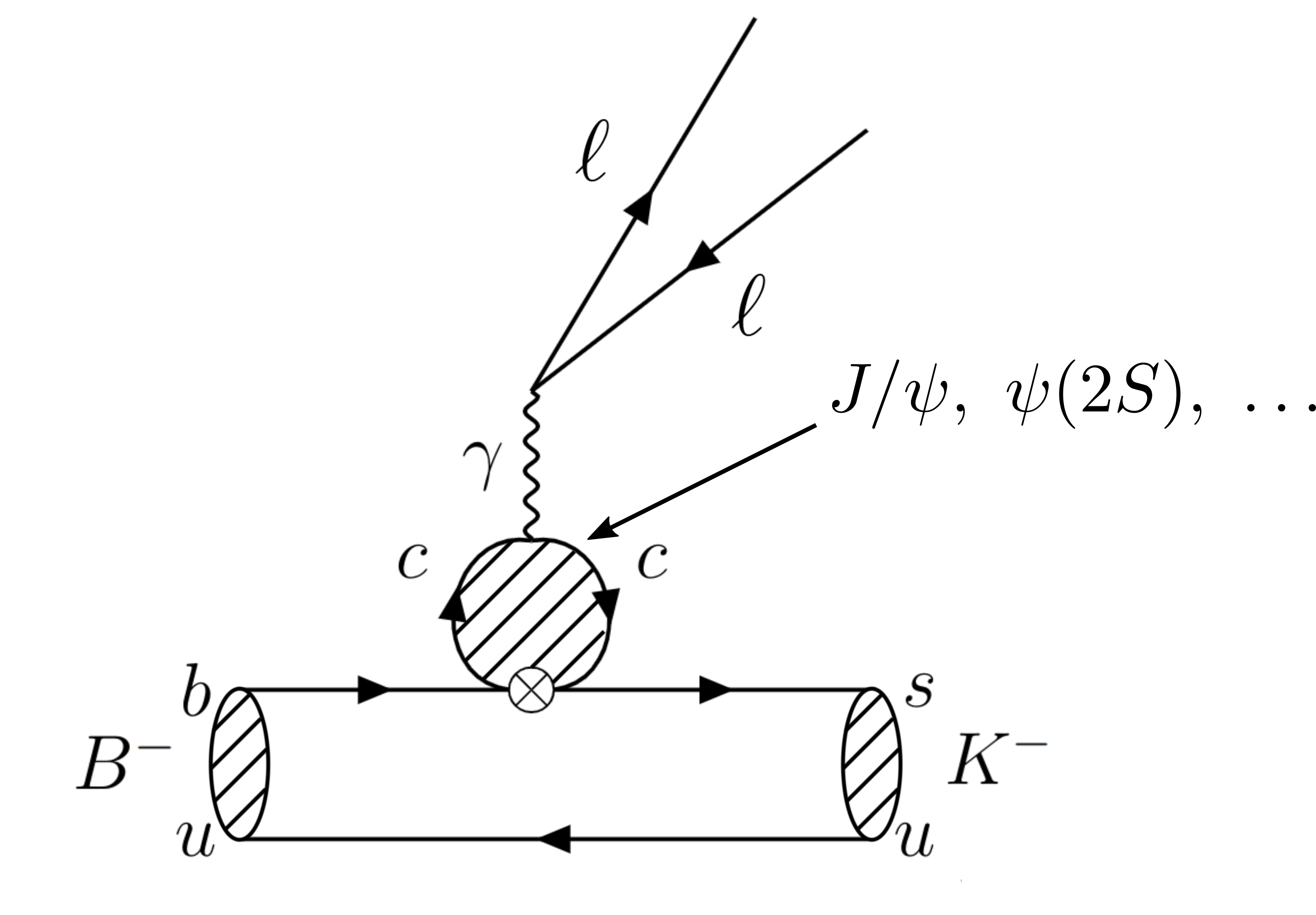}
    \caption{Hadronic long-distance effects in $B^- \to K^-\ell^+\ell^-$.}
    \label{fig:hadronic_LD_effects}
\end{figure}

The decay $B^- \to K^- \ell^+ \ell^-$ can also proceed through an intermediate vector meson $V$, which subsequently decays into the lepton pair. This process originates from the $\mathcal{O}_1^c$ and $\mathcal{O}_2^c$ current-current operators in the low-energy effective Hamiltonian \eqref{eq:ham} through loops and corresponding rescattering effects as illustrated in Fig.~\ref{fig:hadronic_LD_effects}. We note that the hadronic long-distance contributions have been discussed intensively (see e.g.\cite{Khodjamirian:2012rm, Lyon:2013gba, Ciuchini:2015qxb, Capdevila:2017ert,Blake:2017fyh,Chrzaszcz:2018yza, Gubernari:2022hxn}). We account for these effects by replacing $C_9$ by an effective coefficient
\begin{equation}
    C_9^{\rm eff} = C_9 + Y(q^2) \ ,
\end{equation}
where the function $Y \equiv |Y| e^{i \delta_Y}$ with a CP-conserving strong phase $\delta_Y$.

The long-distance contributions are especially large in $q^2$-regions close to a $c \bar c$ resonance peak, where the intermediate meson can go on shell. The large size of non-local effects close to vector resonances is both a blessing and a curse. On the one hand, a larger amplitude means that any direct CP asymmetry will be enhanced near the peaks \cite{Becirevic:2020ssj}. On the other, a large amplitude means that measurements near the resonances are contaminated by large backgrounds. For this reason, experimental studies are usually restricted to regions far away from the peaks \cite{LHCb:2021trn,LHCb:2021lvy}. In this work, we consider $q^2$ bins near the resonances, laying the foundation for future analyses once the challenge of the peaks can be overcome. Since the $Y$ contains non-perturbative physics, we will use experimental data to determine it. We note that the perturbative contribution to $C_9^{\rm eff}$ has been calculated and discussed in \cite{Asatrian:2001de,Asatryan:2001zw,Beneke:2001at,Greub:2008cy,Bobeth:2011gi, Du:2015tda }. Here we do not add this contribution to our description, to avoid possible double counting because we fit to the experimental data (see e.g. discussion in \cite{Huber:2019iqf} for inclusive $B\to X_{s,d}\ell^+\ell^-$ decays).
We adopt a parameterization where the $c\bar{c}$ vector resonances in $B \to K\mu^+\mu^-$ are modelled as a sum of Breit--Wigner distributions following \cite{Kruger:1996cv,Kruger:1996dt}:
\begin{equation}
    Y(q^2) = \sum\limits_j \eta_j e^{i \delta_j} A_j^{\rm res}(q^2) \ ,
    \label{eq:long-distance}
\end{equation}
where $\eta_j$ is the magnitude of a resonance amplitude, $\delta_j$ its CP-conserving phase, and
\begin{equation}
    A_j^{\rm res}(q^2) = \frac{m_{0j} \Gamma_{0j}}{(m_{0j}^2 - q^2) - i m_{0j} \Gamma_j(q^2)} \ , 
\end{equation}
where $\Gamma_j$ and $m_{0j}$ are the width and pole mass of the $j$th resonance, respectively. We use the values of \cite{LHCb:2016due}. The sum runs over the six $c \bar c$ resonances $J/\psi$, $\psi(2S)$, $\psi(3770)$, $\psi(4040)$, $\psi(4160)$, and $\psi(4415)$. In this model, contributions from other broad resonances and continuum contributions are ignored. There also exist resonances involving lighter quarks, but their contribution to the branching ratios is only around $1\%$ \cite{LHCb:2016due}. We neglect these contributions for consistency as we also neglect the doubly Cabibbo-suppressed contributions to the decay amplitude as denoted after \eqref{eq:ham}.

\begin{figure}
    \centering
    \subfloat[$\abs{Y(q^2)}$]{\includegraphics[height=0.25\textwidth]{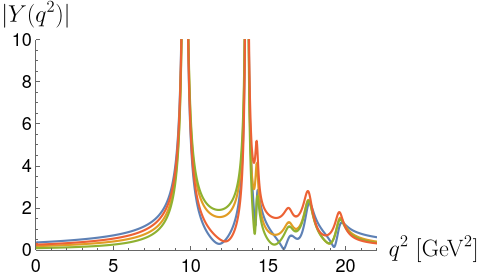}}
    \hfill
    \subfloat[$\delta_Y(q^2)$]{\includegraphics[height=0.25\textwidth]{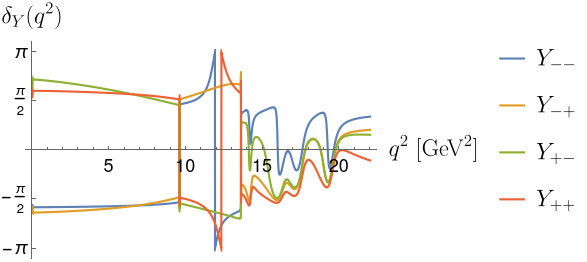}}
    \caption{Absolute value and phase of the hadronic long-distance function $Y(q^2)$. The four colors show the four branches from Ref. \cite{LHCb:2016due} (see text).}
    \label{fig:Y_abs_arg}
\end{figure}

The parameters $\eta$ and the phases $\delta$ can then be determined from experimental data. The LHCb collaboration has measured them by fitting the model to the full invariant mass spectrum \cite{LHCb:2016due}. In the fit four ``branches'' were found, corresponding to different signs for the phases of the $J/\psi$ and $\psi(2S)$ resonances. We denote them by $Y_{--}$, $Y_{-+}$, $Y_{+-}$, and $Y_{++}$, where the first (second) sign matches the sign of the CP-conserving phase of the $J/\psi (\psi(2S))$. Fig. \ref{fig:Y_abs_arg} shows the $|Y(q^2)|$ and $\delta_Y$ for each of the four branches. We can see that the absolute value of $Y(q^2)$ grows largest near the $c \bar c$ resonances, and furthermore that it behaves similarly in all four branches. The overall phase $\delta_Y(q^2)$ differs more: up to $m_{J/\psi}^2 \approx 10 \;\si{\giga\eV^2}$, it is determined almost completely by the phase of the $J/\psi$, so the $Y_{--}, Y_{-+}$ branches are similar, as are the $Y_{+-}, Y_{++}$. Near the $\psi(2S)$ peak, $\delta_Y(q^2)$ is determined by the phase of the $\psi(2S)$. At higher $q^2$, past the two major $c \bar c$ resonances, $\delta_Y(q^2)$ follows a pattern that is similar for all branches. Because the phase $\delta_Y(q^2)$ is distinct for each branch, we can use CP-violating observables that are sensitive to $\delta_Y(q^2)$ to distinguish between the different branches. We will return to this in Section \ref{ch:disentangling_branches}. We note that this description of the hadronic resonances is a model and as such could be significantly improve in the future using either more data or a different parametrization \cite{Blake:2017fyh,Chrzaszcz:2018yza}, see for example the recent discussion in \cite{Isidori:2021tzd}. Nevertheless, it is instructive to consider this model to determine the distinct patterns given by the CP asymmetries.

\subsection{Direct CP violation}
\label{ch:direct_CPV}
In order to generate a direct CP-asymmetry, at least two interfering amplitudes are required, as well as two different CP-violating phases and CP-conserving phases. In the case of $B \to K\mu^+\mu^-$, a CP-violating phase can come from tiny Cabibbo-suppressed interference terms that we neglect, or from NP, manifest as a complex phase in a Wilson coefficient. In our approach, the only source of a CP-conserving phase in this decay is the function $Y(q^2)$ that describes hadronic long-distance effects.

We denote by $\bar \Gamma \equiv \Gamma(B^- \to K^- \ell^+\ell^-)$, $\Gamma \equiv \Gamma(B^+ \to K^+ \ell^+\ell^-)$, and define the differential direct CP asymmetry as
\begin{equation}
\begin{aligned}
    \mathcal{A}_{\rm CP}^{\rm dir}(q^2) \equiv \frac{d\bar \Gamma/dq^2
    - d \Gamma/dq^2}{d\bar \Gamma/dq^2
    + d \Gamma/dq^2}
\end{aligned}
\label{eq:ACP_dir}
\end{equation}
with 
\begin{equation*}
\begin{aligned}
    \frac{d\bar \Gamma/dq^2 - d\Gamma/dq^2}{2}
    &= \frac{4}{3}(1 + 2 \hat m_\ell^2) (\abs{\bar h_V}^2 - \abs{h_V}^2) + 16 \hat m_\ell (\Im[\bar h_V \bar h_{T_t}^*] - \Im[h_V h_{T_t}^*])\\
\end{aligned}
\end{equation*}
\begin{equation}
\begin{aligned}
    = \abs{\mathcal{N}}^2 \frac{\lambda_B}{\sqrt{q^2}} f_+(q^2) \abs{Y(q^2)} \sin \delta_Y(q^2) &\bigg[ \frac{4 (1 + 2 \hat m_\ell^2)}{3 \sqrt{q^2}} f_+(q^2) \abs{C_{9\ell}^{\rm NP}} \sin \phi_9^{\rm NP}\\
    &+ \frac{8 \hat m_\ell}{m_B + m_K} f_T(q^2) \abs{C_{T\ell}^{\rm NP}} \sin \phi_{T\ell}^{\rm NP})\bigg] \ ,
    \label{eq:Adir_numerator}
\end{aligned}
\end{equation}
where the $\bar{h}_i$ were given in \eqref{eq:h_amplitudes_V} and \eqref{eq:h_amplitudes_Tt}. In addition, we have defined the CP-violating phase $\phi_i^{\rm NP}$ of a Wilson coefficient through 
\begin{equation}
    C_{i\ell}^{\rm NP} \equiv \abs{C_{i\ell}^{\rm NP}} e^{i \phi_{i\ell}^{\rm NP}} \ ,
\end{equation}
where we explicitly denote by $\ell$ the lepton flavour. We note that, in order to get direct CP violation in $B \to K \ell^+ \ell^-$, an interference between the hadronic long-distance phase $\delta_Y(q^2)$ and a NP phase $\phi_{9\ell}^{\rm NP}$ or $\phi_{T\ell}^{\rm NP}$ is required. Because the direct CP asymmetry is proportional to $\sin\delta_Y(q^2)$, it inherits the properties of the CP-conserving phase variations as a function of $q^2$ given in Fig.~\ref{fig:Y_abs_arg}a. Consequently, as has been observed in \cite{Becirevic:2020ssj}, the direct CP asymmetry will vary across the $q^2$ spectrum, growing largest close to the first two $c \bar c$ resonance peaks. 

In the SM, tiny amounts of direct CP violation can be generated through interference between the terms proportional to $\lambda_t$ and the doubly Cabibbo-suppressed $\lambda_u$ term which we neglect. For this reason, in the SM $\mathcal{A}_{\rm CP}^{\rm dir}$ is suppressed by a factor $\lambda^2$ relative to \eqref{eq:Adir_numerator}. 
Therefore, an observed sizeable direct CP asymmetry in $B \to K\ell^+\ell^-$ would be a clear sign of new CP-violating physics.

The denominator of the direct CP asymmetry is
\begin{equation}
   \frac{d\bar \Gamma/dq^2 + d \Gamma/dq^2}{2} = \bar{G}_0 + 
G_0
\label{eq:Adir_denominator}
\end{equation}
with $\bar{G}_0$ given in \eqref{eq:G0bar}.

To connect with experiment, we consider the $q^2$-integrated direct CP asymmetry, defined as
\begin{equation}
    \mathcal{A}_{\rm CP}^{\rm dir}[q^2_{\rm min}, q^2_{\rm max}] = \frac{\bar \Gamma[q^2_{\rm min}, q^2_{\rm max}] - \Gamma[q^2_{\rm min}, q^2_{\rm max}]}{\bar \Gamma[q^2_{\rm min}, q^2_{\rm max}] + \Gamma[q^2_{\rm min}, q^2_{\rm max}]} \ ,
    \label{eq:q2_binned_CP_asymm}
\end{equation}
where 
\begin{equation}\label{eq:brdef}
    \Gamma[q^2_{\rm min}, q^2_{\rm max}] = \int_{q^2_{\rm min}}^{q^2_{\rm max}} \frac{d\Gamma}{dq^2} dq^2 \ .
\end{equation}
Note that in \eqref{eq:q2_binned_CP_asymm}, all the differential decay rates have been integrated individually -- this is not the same as integrating \eqref{eq:ACP_dir} over a $q^2$ bin.

The numerator of the $q^2$-binned direct CP asymmetry can be written as \begin{equation}
\begin{aligned}
  \frac{  \bar \Gamma[q^2_{\rm min}, q^2_{\rm max}] - \Gamma[q^2_{\rm min}, q^2_{\rm max}]}{2} &= \rho_{9\rm Im}^\ell \abs{C_{9\ell}^{\rm NP}} \sin \phi_{9\ell}^{\rm NP}
    + \rho_{T\rm Im}^\ell \abs{C_{T\ell}^{\rm NP}} \sin \phi_{T\ell}^{\rm NP} \ ,
    \label{eq:CP_asymm_numerator_numerical}
\end{aligned}
\end{equation}
and the denominator as
\begin{align}
  &\frac{  \bar \Gamma[q^2_{\rm min}, q^2_{\rm max}] + \Gamma[q^2_{\rm min}, q^2_{\rm max}] }{2}=   \Gamma_{\rm SM}^\ell
    + \rho_9^\ell \abs{C_{9\ell}^{\rm NP}}^2
    + \rho_{10}^\ell \abs{C_{10\ell}^{\rm NP}}^2
    + \rho_P^\ell \abs{C_{P\ell}^{\rm NP}}^2
    + \rho_S^\ell \abs{C_{S\ell}^{\rm NP}}^2
    + \rho_T^\ell \abs{C_{T\ell}^{\rm NP}}^2 \nonumber \\
    & \quad\quad\quad\quad + \rho_{9 \rm Re}^\ell \abs{C_{9\ell}^{\rm NP}} \cos \phi_{9\ell}^{\rm NP}
    + \rho_{10\rm Re}^\ell \abs{C_{10\ell}^{\rm NP}} \cos \phi_{10\ell}^{\rm NP}
    + \rho_{P\rm Re}^\ell \abs{C_{P\ell}^{\rm NP}} \cos \phi_{P\ell}^{\rm NP}+ \rho_{T\rm Re}^\ell \abs{C_{T\ell}^{\rm NP}} \cos \phi_{T\ell}^{\rm NP}\nonumber \\
    &  \quad\quad\quad\quad  + \rho_{10P\rm Re}^\ell \abs{C_{10\ell}^{\rm NP}} \abs{C_{P\ell}^{\rm NP}} \cos \phi_{10\ell}^{\rm NP}  \cos \phi_{P\ell}^{\rm NP}
    + \rho_{10P\rm Im}^\ell \abs{C_{10\ell}^{\rm NP}} \abs{C_{P\ell}^{\rm NP}} \sin \phi_{10\ell}^{\rm NP}  \sin \phi_{P\ell}^{\rm NP}\nonumber \\
    & \quad\quad\quad\quad  + \rho_{9T\rm Re}^\ell \abs{C_{9\ell}^{\rm NP}} \abs{C_{T\ell}^{\rm NP}} \cos \phi_{9\ell}^{\rm NP}  \cos \phi_{T\ell}^{\rm NP}
    + \rho_{9T\rm Im}^\ell \abs{C_{9\ell}^{\rm NP}} \abs{C_{T\ell}^{\rm NP}} \sin \phi_{9\ell}^{\rm NP}  \sin \phi_{T\ell}^{\rm NP} \ .
    \label{eq:CP_asymm_denominator_numerical}
\end{align}
The coefficients $\Gamma_{\rm SM}$ and $\rho$ depend on the choice of long-distance model. Numerical values for these coefficients are given in Appendix \ref{ch:coefficients}, with long-distance branches separated where appropriate.

\subsection{Mixing-induced CP violation}
Charged $B$-meson decays exhibit only direct CP violation. Neutral $B^0_q$ mesons show $B^0_q-\bar{B}^0_q$ oscillations, which have profound implications for CP violation. Due to $B^0_q-\bar{B}^0_q$ mixing, interference effects arise between $B^0_q$ and $\bar{B}^0_q$ decaying into the same final state $f$, thereby leading to mixing-induced CP violation. This phenomenon has received a lot of attention in non-leptonic $B$ decays but is also very interesting for the rare modes that we study here. 

The $B^0_d \to K^0 \ell^+ \ell^-$ and $\bar{B}^0_d \to \bar{K}^0 \ell^+ \ell^-$ modes followed by $K^0 \to K^+ \pi^-$ and $\bar{K}^0 \to \pi^+ K^-$, respectively, are flavour specific decays, which can be distinguished through the charges of the subsequent kaon decays. However, if we observe the neutral kaons as $K_S$, both $B^0$ and $\bar{B}^0$ can decay into the same final state. We note that we neglect the CP violation in the neutral kaon system, which appears at the $10^{-3}$ level, and thereby treat $K_S$ as a CP eigenstate. Consequently, for these decays $B^0_q-\bar{B}^0_q$ mixing can generate interference between these decay amplitudes, giving rise to mixing-induced CP violation. Therefore, in the $B^0_d \to K_S\mu^+\mu^-$ channel these interference effects are present. The amplitudes of $B^0_d \to K_S \ell^+ \ell^-$ and $\bar{B}^0_d \to K_S \ell^+ \ell^-$  are related through spectator quarks in the isospin limit to the charged ones of $B^+ \to K^+ \ell^+ \ell^-$ and $B^- \to K^- \ell^+ \ell^-$, respectively, taking the normalisation factor $1/\sqrt{2}$ at the amplitude level coming from the $K_S$ state into account. 

In the presence of $B^0_q-\bar{B}^0_q$ mixing, the time evolution can be expressed as:
\begin{align}\label{eq:decs}
\Gamma(B^0_q \to f) &= \Big[ |g_-(t)|^2 + |\xi_f|^2 |g_+(t)|^2 - 2 \Re \{ \xi_f \  g_+(t) \  g_-(t)^* \} \Big] \Gamma_f, \nonumber \\
\Gamma(\bar{B}^0_q \to f) &= \Big[ |g_+(t)|^2 + |\xi_f|^2 |g_-(t)|^2 - 2 \Re \{ \xi_f \  g_-(t) \  g_+(t)^* \}\Big]{\Gamma}_f,
\end{align} 
where the normalisation $\Gamma_f$ denotes the ``unevolved'' $B_q \to f$ rate and the time dependence enters through: 
\begin{align}
\label{eq:gfun}
g_+(t) \ g_-(t)^* &= \frac{1}{4} \left[ e^{- \Gamma_{L} t} - e^{- \Gamma_{H} t} - 2 i e^{-\Gamma t} \sin{\Delta M t} \right], \nonumber \\
|g_{\mp}(t)|^2 &= \frac{1}{4}  \left[ e^{- \Gamma_{L} t} + e^{- \Gamma_{H} t} \mp 2  e^{-\Gamma t} \cos{\Delta M t} \right].
\end{align}
Here $\Gamma_{H,L}$ is the width of the  ``heavy'' and ``light'' mass eigenstates and $\Delta M = M_H - M_L$ is the mass difference between those eigenstates. The parameter $\xi$ is a  process-specific physical observable that measures the strength of the interference effects: 
\begin{equation}
\xi_f= e^{-i\phi_{q}}\left[e^{\phi_{CP}}\frac{A(\bar{B}^0_q \to f) }{A({B}^0_q \to f)}\right] \ .
\end{equation}
Here $\phi_{CP}$ is a convention-dependent phase, which later cancels in the ratio of the amplitudes, while $\phi_q$ is the $B^0_q-B^0_q$ mixing phase. This phase is sizeable for the $B_d$ system and takes the value \cite{Barel:2020jvf,Barel:2022wfr}
\begin{equation}
    \phi_d = (44.4^{+1.6}_{-1.5})^\circ \ ,
    \label{phid}
\end{equation}
which is extracted from $B_d^0\to J/\psi K_S$ and which also takes penguin contributions into account. We note here that the mixing phase $\phi_s$ in the $B_s$ system, e.g. in the $B_s^0 \to f_0 \ell^+ \ell^-$ decay \cite{Descotes-Genon:2020tnz}, is much smaller, for which reason $B_s^0$ decays are more sensitive to NP effects.
We can then introduce the time-dependent decay rate asymmetry (following the notation in \cite{Fleischer:2017yox}):
\begin{equation}
\begin{aligned}
    \frac{\Gamma(B^0_q(t) \to f) 
    - \Gamma(\bar B^0_q(t) \to f)}{\Gamma(B^0_q(t) \to f)
    + \Gamma(\bar B^0_q(t) \to f)} 
    &=\frac{\mathcal{C}\cos(\Delta M t)  +\mathcal{S} \sin (\Delta M t) }{\cosh (\frac{\Delta \Gamma_q}{2} t) + \mathcal{A}_{\Delta \Gamma} \sinh (\frac{\Delta \Gamma_q}{2} t)} \ ,
\end{aligned}
\label{eq:time_dependent_CP_asymmetry}
\end{equation}
where $\Delta \Gamma_q \equiv \Gamma_L-\Gamma_H$ denotes the decay width difference in the $B_q$ system. We note that $\Delta \Gamma_q$ is sizeable in the $B_s$ system but negligible in the $B_d$, therefore making the observable $\mathcal{A}_{\Delta \Gamma}$ not accessible in $B^0_d \to K_S \ell^+\ell^-$. A full angular analysis would give rise to 6 angular observables as discussed in \cite{Descotes-Genon:2020tnz}. As the time-dependent measurement of these observables appears to be challenging, our starting point is to consider rates integrated over the full range $\theta$ (as in the charged case above), such that only the $S$-wave component remains. At a decay time $t=0$, where mixing effects are switched off, the untagged differential decay rate then yields:
\begin{align}
\frac{d\Gamma(B_d\to K_S\ell^+\ell^-)\pm d\Gamma(\bar{B}_d\to K_S\ell^+\ell^-)}
{ds}
&=2 [G_0 \pm \bar{G}_0]  \ ,
\end{align}
with $\bar{G}_0$ given in \eqref{eq:G0bar} but with the replacements
\begin{equation}\label{eq:hneutral}
    \bar{h}_X(\bar{B}_d\to K_S \ell^+\ell^-) = \frac{1}{\sqrt{2}} \bar{h}_X(B^-\to K^-\ell^+\ell^-) \ , \quad  {h}_X(\bar{B}_d\to K_S \ell^+\ell^-) = \frac{1}{\sqrt{2}} {h}_X(B^-\to K^-\ell^+\ell^-) \ , 
\end{equation}
to take the normalization of the $K_S$ into account. 

From \eqref{eq:decs} and \eqref{eq:gfun}, we find 
\begin{align}\label{eq:J+Jt}
G_0(t)-\bar G_0(t) &=e^{-\Gamma t}\Big[(G_0 - \bar G_0)\cos(x\Gamma t) - s_0 \sin(x\Gamma t)\Big]\ , \\
G_0(t)+\bar G_0(t) &=e^{-\Gamma t}\Big[(G_0 + \bar G_0)\cosh(y\Gamma t) - h_0 \sinh(y\Gamma t)\Big]\ ,
\label{eq:Gs}
\end{align}
where $y = \Delta \Gamma/(2 \Gamma)$ and $x= \Delta M/ \Gamma$. And \cite{Descotes-Genon:2020tnz} 
\begin{equation}
\begin{aligned}
    s_0 = 2 \Im \bigg[ e^{-i \phi_d} &\bigg( \frac{4}{3} (1 + 2 \hat m_\ell^2) \Tilde h_V h_V^* + \frac{4}{3} \beta_\ell^2 \Tilde h_A h_A^* + 2 \beta_\ell^2 \Tilde h_S h_S^* + 2 \Tilde h_P h_P^*\\
    &+ \frac{8}{3} (1 + 8 \hat m_\ell^2) \Tilde h_{T_t} h_{T_t}^* + \frac{4}{3} \beta_\ell^2 \Tilde h_T h_T^* \bigg) \bigg] - 16 \hat m_\ell \Re \bigg[ e^{- i \phi_d} \Tilde h_V h_{T_t}^* - e^{i \phi_d} h_V \Tilde h_{T_t}^* \bigg] \ ,
    \label{eq:s0}
\end{aligned}
\end{equation}
The normalization is as in \eqref{eq:hneutral} and $\tilde{h}_X= \eta_X \bar{h}_X$, with for the $K_S$, $\eta_V=\eta_A=\eta_P=\eta_{T_t}=-1$ and $\eta_S=\eta_T=1$ (see \cite{Descotes-Genon:2020tnz, Descotes-Genon:2015hea} for a detailed discussion). The coefficient $h_0$ is also given in \cite{Descotes-Genon:2020tnz}. 

Finally, we can write the CP asymmetries as\footnote{We note that we have $\mathcal{A}_{\rm CP}^{\rm mix} = 2\sigma_0$, where $\sigma_0$ was used in \cite{Descotes-Genon:2020tnz}.}
\begin{align}
   \mathcal{C} & \equiv \frac{G_0 - \bar{G}_0}{G_0+\bar{G}_0} = - \mathcal{A}_{\rm CP}^{\rm dir} (B^-\to K^- \ell^+ \ell^-) \ ,\\
   \mathcal{S}&\equiv \frac{-s_0}{G_0+\bar{G}_0} \equiv -  \mathcal{A}_{\rm CP}^{\rm mix} \ , \label{eq:Sdef} 
\end{align}
where the charged direct CP asymmetry was already introduced in \eqref{eq:ACP_dir}.

Unlike the direct CP asymmetry, which is sensitive only to a CP-violating phase in $C_9$ or $C_T$, the mixing-induced CP asymmetry is affected by complex phases in any of the Wilson coefficients. The mixing phase $\phi_d$, which is experimentally found to be sizeable, plays a key role, leading to a large mixing-induced CP asymmetry. Therefore, having new complex Wilson coefficients involved, this CP asymmetry can be shifted away from its SM value. We provide the SM prediction in the next section.

\subsection{Standard Model predictions and input parameters}
For the SM Wilson coefficients, we use \cite{Descotes-Genon:2013vna}
\begin{equation}
    C_7^{\rm SM} = -0.292, \quad C_9^{\rm SM} = 4.07, \quad C_{10}^{\rm SM} = -4.31 \ ,
\end{equation}
which are flavour universal (i.e.\ equal for $\ell=\mu=e$) and determined at $\mu=m_b$. In addition, we use the quark masses in ${\overline{\rm MS}}$ at $m_b$: $\bar{m}_b(m_b) = 4.18\pm 0.03$ and $\bar{m}_s(m_b)= 0.078\pm 0.007$ \cite{ParticleDataGroup:2020ssz}. For the local form factors in \eqref{eq:localff}, we use the recent lattice QCD determination of \cite{Parrott:2022rgu}. 

To determine the CKM elements $V_{ts}V_{tb}^*$, we exploit CKM unitarity to write (see \cite{DeBruyn:2022zhw}): 
\begin{equation}\label{eq:vtsvtb}
    V_{ts}V_{tb}^* = -V_{cb} \bigg[1 - \frac{\lambda^2}{2}(1-2 \bar{\rho} + 2i \bar{\eta}) \bigg] + \mathcal{O}(\lambda^6)  \ ,
\end{equation}
which introduces the dependence of the apex of the Unitarity Triangle (UT) through $\bar{\rho}$ and $\bar{\eta}$ at $\mathcal{O}(\lambda^2)$ with respect to the leading $|V_{cb}|$ contribution. For completeness, we do include here the $\lambda^2$-suppressed terms such that we have an imaginary part arising from the CKM matrix elements.

The discrepancies between the inclusive and exclusive $|V_{ub}|$ and $|V_{cb}|$ determinations lead to different pictures for the allowed parameter spaces for NP in  $B_q$-meson mixing. We employ the recent results from \cite{DeBruyn:2022zhw} where a CKM fit was performed using both the inclusive and exclusive case. On top of these, a hybrid approach with the exclusive $|V_{ub}|$ and inclusive $|V_{cb}|$ values was studied \cite{DeBruyn:2022zhw}. 
We note that for determining the CKM factor in \eqref{eq:vtsvtb}, the difference for $|V_{ub}|$ is marginal and only enters via higher order corrections, which we neglect. In this limit, the hybrid scenario coincides with the inclusive one. We provide our results separately, first for the inclusive/hybrid and then for the exclusive case. 

For $|V_{cb}|$, we take the result from inclusive $b\to c\ell\nu$ decays \cite{Bordone:2021oof}:
\begin{equation}
    |V_{cb}|_{\rm incl} = (42.16 \pm 0.50)\times 10^{-3} \ ,
\end{equation}
which agrees with the recent determination in \cite{Bernlochner:2022ucr}. Finally,  \eqref{eq:vtsvtb} then gives 
\begin{equation}\label{eq:hybrid}
V_{ts}V_{tb}^*|_{\rm incl/hybrid} = (-41.4\pm 0.5 + 0.7 i)\times 10^{-3}, \quad |V_{ts}V_{tb}^*|_{\rm incl/hybrid} = (41.4 \pm 0.5)\times 10^{-3} \ ,
\end{equation}
where here and in the following we take into account the tiny imaginary part from the CKM elements. Taking the exclusive $|V_{cb}|$ given in \cite{DeBruyn:2022zhw} from \cite{HFLAV:2019otj}, we find 
\begin{equation}
|V_{ts}V_{tb}^*|_{\rm excl} = (38.4 \pm 0.5)\times 10^{-3},
\end{equation}
which differs from  \eqref{eq:hybrid} at the $4\sigma$ level to the different $|V_{cb}|$ values.

Finally, we find for the branching ratio in the $1.1 \;{\rm GeV}^2 <q^2 < 6.0 \; {\rm GeV}^2$ range:
\begin{equation}
    \mathcal{B}(B^- \to K^- \mu^+\mu^-)^{\rm SM}[1.1,6.0]_{\rm incl/hybrid} = (1.83 \pm 0.14) \times 10^{-7} \ ,
    \label{eq:SM_BR_lowq2}
\end{equation}
where we have added the variation of the different long distance branches as an additional uncertainty. Even when including those, the uncertainty is still dominated by the form factor uncertainty. We note that we do not include a systematic uncertainty from our long-distance assumptions. Comparing this with the experimental measurement of the branching ratio in this $q^2$-bin \cite{LHCb:2014cxe} \footnote{We have converted the differential rate given in \cite{LHCb:2014cxe} by multiplying with the appropriate bin-width.}:
\begin{equation}
    \mathcal{B}(B^\pm \to K^\pm \mu^+\mu^-) [1.1, 6.0] = (1.19 \pm 0.07)  \times 10^{-7} \ , 
\end{equation}
we find
\begin{equation}
\frac{   \mathcal{B}(B^- \to K^- \mu^+\mu^-)^{\rm SM}[1.1,6.0]_{\rm {{incl/hybrid}}}}{\mathcal{B}(B^\pm \to K^\pm \mu^+\mu^-) [1.1, 6.0]} -1 = 0.54 \pm 0.15 \ , 
\end{equation}
which is a $3.5\sigma$ deviation from zero. Taking the exclusive $|V_{cb}|$ determination, we find 
\begin{equation}
    \mathcal{B}(B^- \to K^- \mu^+\mu^-)^{\rm SM}[1.1,6.0]_{\rm excl} = (1.57 \pm 0.13) \times 10^{-7} \ ,
    \label{eq:SM_BR_lowq2_excl}
\end{equation}
and
\begin{equation}
    \frac{   \mathcal{B}(B^- \to K^- \mu^+\mu^-)^{\rm SM}[1.1,6.0]_{\rm excl}}{\mathcal{B}(B^\pm \to K^\pm \mu^+\mu^-) [1.1, 6.0]} -1 = 0.33 \pm 0.13 \ ,
\end{equation}
which lies $2.4\sigma$ from the SM. This study shows that the unsatisfactory determination of the CKM factors has a profound impact on the comparison between the data and the SM prediction for the corresponding branching ratio, thus it has to be resolved.

Finally, we note that our determination is based on our assumptions for the hadronic long-distance effects. As such, our prediction differs slightly from \cite{Parrott_SM_predictions} which found:
\begin{equation}
    \mathcal{B}(B^- \to K^- \mu^+\mu^-)^{\rm SM}[1.1,6.0] = (2.07 \pm 0.17) \times 10^{-7} \ ,
\end{equation}
which shows a $4.7\sigma$ deviation from the SM. In this case, the authors used a different model for the $Y(q^2)$ including the perturbative contribution discussed previously. Their variation of the charm mass results in a larger uncertainty than ours. The SM predicitions for these modes, including long-distance effects, have been discussed in a long range of papers, most recently in \cite{Gubernari:2022hxn}. They also find a significant tension with the SM.

In addition, we give the prediction for the branching ratio in the $7 \;{\rm GeV}^2 <q^2 < 8\; {\rm GeV}^2$ range:
\begin{equation}
    \mathcal{B}(B^+ \to K^+ \mu^+\mu^-)^{\rm SM}[7,8] = (0.367 \pm 0.031) \times 10^{-7} \ .
\label{eq:theoBR_7to8}
\end{equation}

Finally, we note that for electrons, the branching ratios are almost identical, differing only at the third digit through phase-space effects.

Using the value of $\phi_d$ in  \eqref{phid}, we determine the values of the CP asymmetries. Since we neglect the $\lambda^2$ suppressed $\lambda_u$ contributions, we have 
\begin{equation}
\mathcal{A}_{\rm CP}^{\rm dir}|_{\rm SM} = 0,
\end{equation}
for all $q^2$ regions. 
The mixing-induced CP asymmetry for  $1.1 \;{\rm GeV}^2 <q^2 < 6\; {\rm GeV}^2$:
\begin{equation}
    \mathcal{A}_{\rm CP}^{\rm mix}|_{\rm SM} = 0.72 \pm 0.02  ,
\end{equation}
which is exactly equal to $\phi_d$ if the tiny CKM phase is neglected \cite{Descotes-Genon:2020tnz}. The tiny uncertainty stems thus mainly from $\phi_d$. In addition, the CKM factors drop out and we are not affected by the difference in $|V_{cb}|$ as for the branching ratio.

\section{\boldmath Fingerprinting New Physics with direct and mixing-induced CP asymmetries in $B \to K\mu^+\mu^-$}
\label{ch:correlations}

\subsection{Experimental constraints}
\label{ch:exp_bounds}
\begin{figure}[t]
    \centering
    \includegraphics[width=0.6\textwidth]{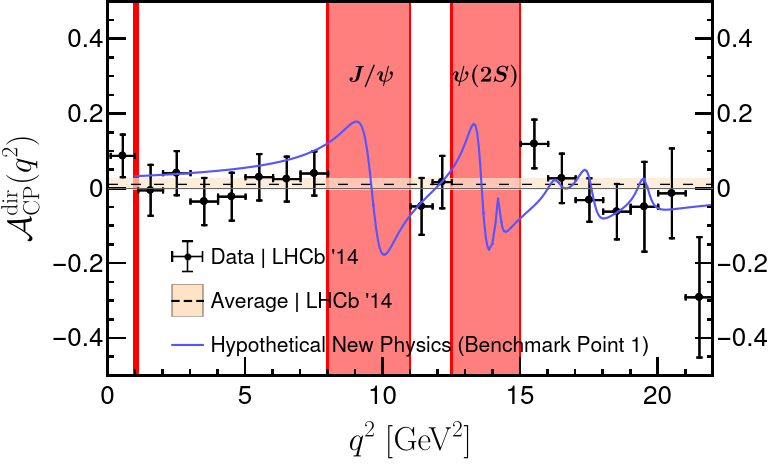}
    \caption{Illustration of the experimental measurements of the direct CP asymmetry of $B^- \to K^- \mu^+ \mu^-$ by the LHCb collaboration \cite{LHCb:2014mit}. The blue line shows the prediction for benchmark point 1 given in~\eqref{eq:staranddiamond} (see also \cite{Becirevic:2020ssj}).}
    \label{fig:ACP_bounds}
\end{figure}

We consider data on the branching ratio \cite{LHCb:2014cxe} and the direct CP asymmetry \cite{LHCb:2014mit} of the $B^+ \to K^+ \mu^+\mu^-$ channel in the bin of $q^2 \in [7,8] \:\si{\giga\eV^2}$, where the LHCb collaboration finds \cite{LHCb:2014cxe} \begin{equation}
    \mathcal{B}(B^+ \to K^+ \mu^+\mu^-)[7,8] = (23.1 \pm 1.8) \times 10^{-9}
\label{eq:expBR_7to8}
\end{equation}
and \cite{LHCb:2014mit}
\begin{equation}\label{eq:adirmeas}
        \mathcal{A}_{\rm CP}^{\rm dir}[7,8] = 0.041 \pm 0.059 \ .
\end{equation}
Here we have added the statistical and systematic uncertainties in quadrature. In this paper, we focus on this $q^2$-bin because it lies close to the $J/\psi$ resonance at $m_{J/\psi}^2 = \SI{9.6}{\giga\eV^2}$, where the direct CP asymmetry could be enhanced \cite{Becirevic:2020ssj}. In Fig.~\ref{fig:ACP_bounds}, we depict the experimental data in the different $q^2$-bins as presented in \cite{LHCb:2014mit}. The average of over all the bins is \cite{LHCb:2014mit}
\begin{equation}
    \mathcal{A}_{\rm CP}^{\rm dir} = 0.012 \pm 0.017 \ .
\end{equation}

We consider three scenarios:
\begin{equation}
\begin{aligned}
    &\text{Scenario 1:} \quad C_{9\mu}^{\rm NP} \neq 0 \ ,\\
    &\text{Scenario 2:} \quad C_{9\mu}^{\rm NP} = -C_{10\mu}^{\rm NP} \neq 0 \ ,\\
    &\text{Scenario 3:} \quad C_{10\mu}^{\rm NP} \neq 0 \ .
\end{aligned}
\label{eq:NP_ranges}
\end{equation}
All three scenarios fit the data better than the SM with pulls of $(4$-$7)\sigma$ \cite{Altmannshofer:2021qrr,Gubernari:2022hxn,Geng:2021nhg,Carvunis:2021jga,Mahmoudi:2022hzx,SinghChundawat:2022zdf}.

\begin{figure}[t]
    \centering
    \subfloat[$C_{9\mu}^{\rm NP}$ only]{\includegraphics[width=0.3\textwidth]{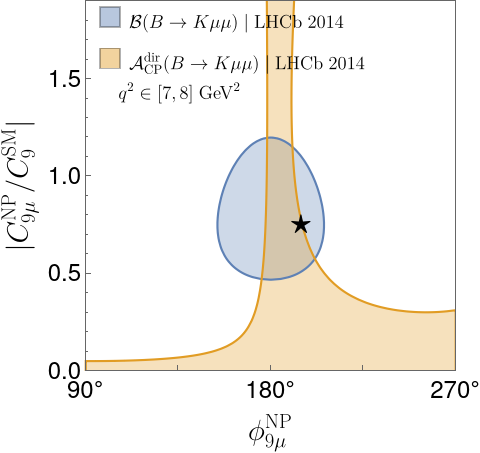}}
    \hfill
    \subfloat[$C_{9\mu}^{\rm NP} = -C_{10\mu}^{\rm NP}$]{\includegraphics[width=0.3\textwidth]{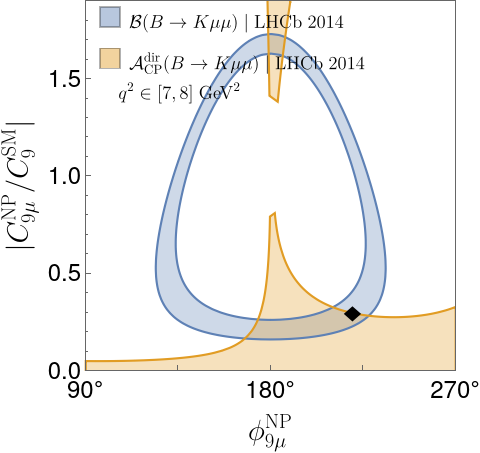}}
    \hfill
   \subfloat[$C_{10\mu}^{\rm NP}$ only]{\includegraphics[width=0.3\textwidth]{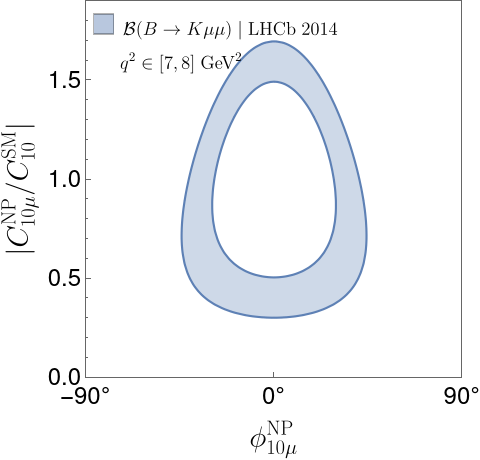}}
    \caption{Experimental $1\sigma$ constraints on the three NP scenarios specified in~\eqref{eq:NP_ranges}. The star and diamond indicate the benchmark points 1 and 2 given in \eqref{eq:staranddiamond}, respectively. 
    }
    \label{fig:C9mu_bounds_scenario_1}
\end{figure}

In Fig. \ref{fig:C9mu_bounds_scenario_1}, we illustrate the $1\sigma$-experimental constraints on the complex $C_{9\mu}$ and $C_{10\mu}$ coefficients for the three NP scenarios in \eqref{eq:NP_ranges}. We note that for Scenario 3, shown in Fig.~\ref{fig:C9mu_bounds_scenario_1}c, there is no bound from the direct CP asymmetry, as can be seen also from \eqref{eq:Adir_numerator}. For simplicity, we show results using the $Y_{--}(q^2$) hadronic long-distance branch. In addition, we use the hybrid scenario for the CKM factors. For the other three branches, the constraints from the branching ratio remain similar while the allowed region from the direct CP asymmetry changes. This dependence on the direct CP asymmetry to the choice of the branch is an important feature that we will exploit in the remainder of this paper. 
In Fig.~\ref{fig:C9mu_bounds_scenario_1}, we have indicated two benchmark points with large CP-violating phases, allowed by the current data: 
\begin{equation}
\begin{aligned}
    \text{Benchmark Point 1:}& \quad \abs{C_{9\mu}^{\rm NP}}/\abs{C_9^{\rm SM}} = 0.75 \ , \quad\quad  \phi_{9\mu}^{\rm NP} = 195^\circ \ ,\\
    \text{Benchmark Point 2:}& \quad \abs{C_{9\mu}^{\rm NP}}/\abs{C_9^{\rm SM}} = \abs{C_{10\mu}^{\rm NP}}/\abs{C_9^{\rm SM}}= 0.30\quad \phi_{9\mu}^{\rm NP} = \phi_{10\mu}^{\rm NP} - \pi = 220^\circ \ .
\end{aligned}
\label{eq:staranddiamond}
\end{equation}
We will use these parameter sets for illustrative purposes in Sec.~\ref{ch:extracting_WCs}. In Fig.~\ref{fig:ACP_bounds}, we show benchmark scenario 1 to demonstrate the enhancement of the direct CP asymmetry in the resonance region. A similar scenario was considered in \cite{Becirevic:2020ssj}. 

In \eqref{eq:NP_ranges}, we only consider scenarios with $C_{9\mu}$ and/or $C_{10\mu}$. We do not discuss NP entering through $C_{S\mu}$ and $C_{P\mu}$, because any significant contribution of these coefficients to $B\to K\ell^+\ell^-$ would have a much larger, clearly visible influence on $B_s^0 \to \mu^+\mu^-$ \cite{Fleischer:2017ltw} as they would lift the helicity suppression. In addition, we do not explicitly consider tensor couplings \cite{Beaujean:2015gba}. If only $C_{T\mu}$ were present, the $B \to K\mu^+\mu^-$ rate in     \eqref{eq:CP_asymm_denominator_numerical} reduces to
\begin{equation}
\Gamma[1.1,6.0] =  \Gamma_{\rm SM} +  \rho_T^\ell \abs{C_{T\ell}^{\rm NP}}^2 + \rho_{T\rm Re}^\ell \abs{C_{T\ell}^{\rm NP}} \cos \phi_{T\ell}^{\rm NP}\ , 
\end{equation}
where the $\rho$ are given in Appendix~\ref{ch:coefficients} from which we find $\rho_{T\rm Re}<\rho_T$. Therefore, the branching ratio will always be pushed upwards, independent of the phase difference $\phi_T$. This situation changes if also NP in other Wilson coefficients is present, in which case the lower value of the branching ratio can be accounted for (see \eqref{eq:CP_asymm_denominator_numerical}). 
For the direct CP asymmetry, we explicitly find
\begin{equation}
    \mathcal{A}_{\rm CP}^{\rm dir} = \frac{\rho_{T\rm Im}^\ell |C_{T\ell}^{\rm NP}| \sin\phi_{T\ell}^{\rm NP}}{\Gamma_{\rm SM}^\ell+  {\rho_T^\ell} |C_{T\ell}^{\rm NP}|^2} \ .
\end{equation}
We note that in this case $ \mathcal{A}_{\rm CP}^{\rm dir} < 0.03$, because $\rho_{T}^\ell >> \rho_{T\rm Im}^\ell$. Therefore, we focus on NP in $C_9$ and $C_{10}$.

\subsection{Correlations between CP-violating observables}
\label{ch:correlations_one_branch}
\begin{figure}
    \centering
    \includegraphics[width=0.7\textwidth]{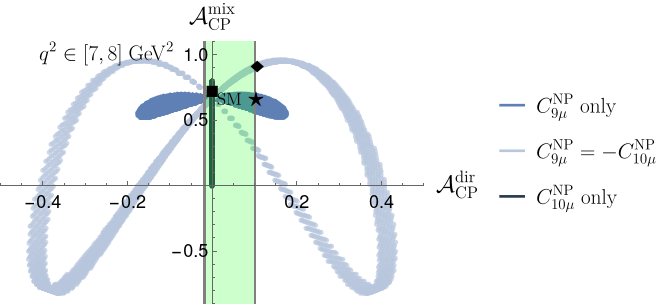}
    \caption{Correlations between the CP asymmetries $\mathcal{A}_{\rm CP}^{\rm dir}$ and $\mathcal{A}_{\rm CP}^{\rm mix}$ for the three NP scenarios in \eqref{eq:NP_ranges}. The SM point is marked by a square, while the current experimental bound on the direct CP asymmetry is illustrated by a green vertical band.
      }
\label{fig:Adir_Amix_correlation_toy_scenario}
\end{figure}

We will now demonstrate how to distinguish between the three NP scenarios in \eqref{eq:NP_ranges} using the correlations between the direct and mixing-induced CP asymmetries of $B \to K\mu^+\mu^-$. To illustrate the method, we first consider only the $Y_{--}$ branch. Considering now NP Wilson coefficients that are within $1\sigma$ of the branching ratio measurement in \eqref{eq:expBR_7to8}, while adding experimental and theoretical uncertainties in quadrature, gives the correlations in \ref{fig:Adir_Amix_correlation_toy_scenario}. In addition, we show the experimental constraint on the direct CP asymmetry for the $B^+\to K^+$ decay given in \eqref{eq:adirmeas}. We observe that each scenario leaves a distinct ``fingerprint'' in the $\mathcal{A}_{\rm CP}^{\rm dir}$-$\mathcal{A}_{\rm CP}^{\rm mix}$ plane. For Scenario 1, with only $C_{9\mu}^{\rm NP}$, the direct CP asymmetry can range from $[-0.2, 0.2]$, while the mixing-induced CP asymmetry remains close to the SM prediction. On the other hand, Scenario 2 results in mixing-induced CP asymmetries as large as $\pm 0.4$, although parts of the allowed region are already in tension with the measurements of the direct CP asymmetries. With a complex phase 
in $C_{10\mu}$ only (Scenario 3), the direct CP asymmetry remains zero. Nevertheless, it allows for large mixing-induced CP asymmetries.

These studies illustrate the power of the correlations between the CP-violating observables to distinguish different NP scenarios.  

\subsection{Distinguishing hadronic long-distance branches}
\label{ch:disentangling_branches}
\begin{figure}
    \centering
    \subfloat[${q^2 \in [7,8] \; \si{\giga \eV^2}}$]{\includegraphics[width=0.9\textwidth]{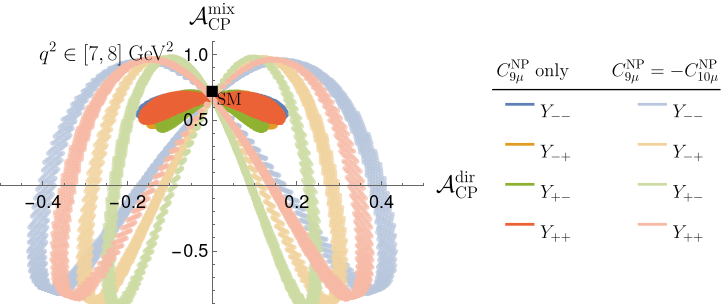}}\\
    \hfill
    \subfloat[${q^2 \in [11,11.8] \; \si{\giga \eV^2}}$]{\includegraphics[width=0.49\textwidth]{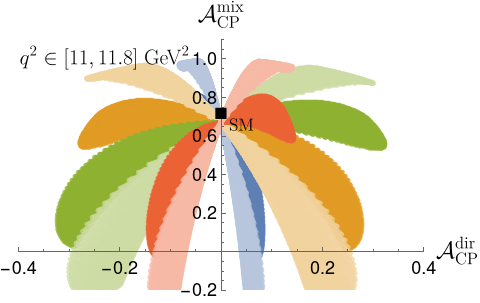}}
    \hfill
    \subfloat[${q^2 \in [14,15] \; \si{\giga \eV^2}}$]{ \includegraphics[width=0.49\textwidth]{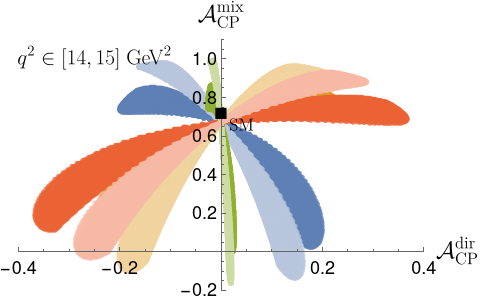}}
    \caption{Correlations between $\mathcal{A}_{\rm CP}^{\rm dir}$ and $\mathcal{A}_{\rm CP}^{\rm mix}$ in different $q^2$ bins including all four hadronic long-distance branches.}
    \label{fig:correlation_all_branches}
\end{figure}

In Fig.~\ref{fig:correlation_all_branches}a, we show the correlations between the CP asymmetries for the $Y_{--}, Y_{-+}, Y_{+-}$ and $Y_{++}$ branches for Scenario 1 and 2 again in the range of $q^2 \in [7,8]$ GeV$^2$. We note that Scenario 3, with only $C_{10\mu}^{\rm NP}$, is not included as it is not influenced by the different branches. The branch multiplicity makes it more challenging to disentangle the different NP scenarios. 

In order to distinguish the branches, we use the high sensitivity of the direct CP asymmetry to the long-distance model and the strong dependence on the $q^2$-bin. Exploiting these features, we can once again distinguish between the branches by using the CP asymmetry correlations in different parts of the $q^2$-spectrum. In Fig. \ref{fig:correlation_all_branches}b and \ref{fig:correlation_all_branches}c, we show the correlations for two bins with $q^2>11$ GeV$^2$. These correlations are obtained by varying the Wilson coefficients as follows: 
\begin{itemize}
\item Scenario 1: $\abs{C_{9\mu}^{\rm NP}}/\abs{C_9^{\rm SM}} \in [0, 0.75] $, $\phi_{9\mu}^{\rm NP} \in [90,270]^\circ$. 
\item Scenario 2: $\abs{C_{9\mu}^{\rm NP}}/\abs{C_9^{\rm SM}} \in [0, 0.50]$, $\phi_{9\mu}^{\rm NP} \in [90,270]^\circ$.
\end{itemize}

From \ref{fig:correlation_all_branches}, we observe that the regions that overlap at lower $q^2$ drift apart. To understand this, we consider the sign of the direct CP asymmetry:
\begin{equation}
    {\rm sign}[\mathcal{A}_{\rm CP}^{\rm dir}] = \sin\phi_{9\mu}^{\rm NP} \times \sin\delta_Y(q^2) \ .
\end{equation}
In the bin $q^2 \in [1.1,6.0] \;\si{\giga\eV^2}$, the $J/\psi$ resonance provides the dominant contribution to $\delta_Y(q^2)$. For $q^2 \in [12.5,13.5] \; \si{\giga \eV^2}$, the $\psi(2S)$-threshold opens and drives the sign of $\delta_Y(q^2)$ (see Fig. \ref{fig:Y_abs_arg}b). At even higher $q^2$, the higher $c \bar c$ resonances dominate and we are no longer sensitive to the phases of the first two resonances. Due to all the different contributions to $\delta_Y$ at different $q^2$, the sign of the direct CP asymmetry varies across the $q^2$ spectrum, thereby giving distinct fingerprints. We have specifically picked the $q^2$-bins of Fig. \ref{fig:correlation_all_branches} to illustrate how branches that are close together in some $q^2$-bins separate in other bins. 

By measuring the CP asymmetries in different parts of the $q^2$ spectrum, we can thus differentiate between the hadronic long-distance branches.

\subsection{Extracting Wilson coefficients from the CP asymmetries}
\label{ch:extracting_WCs}
If the CP asymmetries in $B \to  K\mu^+\mu^-$ deviate from their SM values, the next goal is to extract the Wilson coefficients. In the general scenario of complex and independent NP contributions to both $C_{9\mu}$ and $C_{10\mu}$, we have four parameters entering the observables and therefore we need at least four observables to determine them from the measured data. Here we demonstrate a minimal scenario with four observables, but stress that also other information from different bins could be used. To this end, we use the direct and mixing-induced CP asymmetries and the CP-averaged integrated branching ratio of $B \to K \mu^+\mu^-$ in two different $q^2$ bins. Utilizing the strength of each observable, we consider:
\begin{itemize}
\item the direct CP asymmetry \eqref{eq:q2_binned_CP_asymm} integrated over $q^2 \in [8,9]$,
\item the mixing-induced CP asymmetry \eqref{eq:Sdef} in $q^2 \in [1.1,6.0]$,
\item  the branching ratio in $q^2 \in [1.1,6.0]$ and $q^2 \in [15,22]$. 
\end{itemize}
The branching ratios and the mixing-induced CP asymmetry are considered outside the resonance region and thus less affected by long-distance effects. On the other hand, for the direct CP asymmetry we pick the bin close to the $J/\psi$ peak at $\SI{9}{\giga \eV^2}$ where it may be enhanced. 

\begin{figure}[t]
    \centering
    \includegraphics[width=0.46\textwidth]{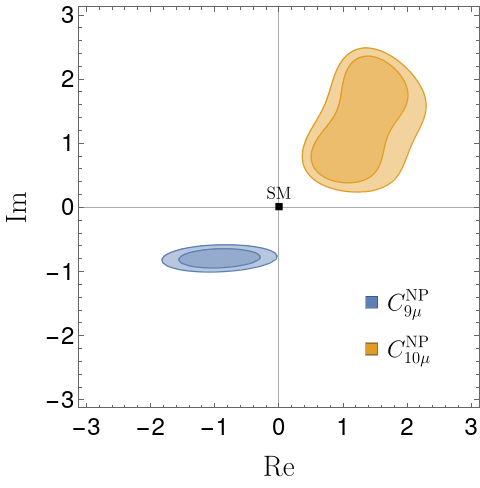}
    \caption{Constraints on $C_{9\mu}$ and $C_{10\mu}$ in the complex plane assuming the input measurements and uncertainties given in ~\eqref{eq:input}. The lines show the $68\%$ and $90\%$ C.L. contours.}
    \label{fig:fitresults}
\end{figure}
\begin{figure}[t]
    \centering
    \includegraphics[width=0.48\textwidth]{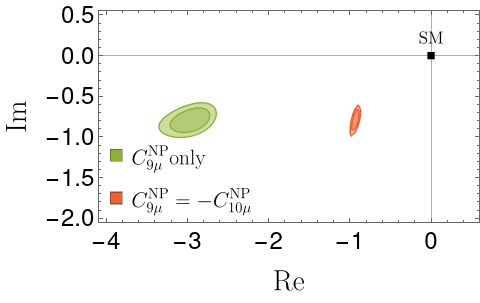}
    \caption{Constraints in the complex plane on $C_{9\mu}$, assuming $C_{9\mu}$ only and $C_{9\mu} = -C_{10\mu}$ for the uncertainties as specified by \eqref{eq:input} and 
    \eqref{eq:input_scenario1}.}
    \label{fig:fitresultscons}
\end{figure}
In order to illustrate the extraction of Wilson coefficients from these four observables, we consider benchmark point 2 specified in~\eqref{eq:staranddiamond}. This NP scenario corresponds to 
\begin{align}\label{eq:input}
\mathcal{A}_{\rm CP}^{\rm dir}[8,9] &= 0.16 \pm 0.02\ , & \mathcal{A}_{\rm CP}^{\rm mix}[1.1,6] &= 0.94 \pm 0.04 \ ,  \\
\mathcal{B}[1.1,6.0] &= (1.15\pm 0.02)\times 10^{-7} \ , &  \mathcal{B}[15,22] &= (0.908\pm 0.018)\times 10^{-7} \nonumber \ ,
\end{align}
where we consider a possible future scenario for the experimental uncertainties. The uncertainties on our inputs in \eqref{eq:input} allows us to explore how the precision on the CP asymmetries translates into allowed regions for the Wilson coefficients. Due to the complexity of the system of equations, we perform a $\chi^2$ fit\footnote{If NP only enters through $C_{9\mu}$ or $C_{9\mu} = -C_{10\mu}$, we can solve the system analytically using only the direct and mixing-induced CP asymmetries as demonstrated in Appendix \ref{ch:analytical_solution}.} while setting theory uncertainties to zero. Figure!\ref{fig:fitresults} shows $68
\%$ and $90\%$ C.L. regions for the extracted Wilson coefficients in the complex plane. We find a good determination of the imaginary part of $C_{9\mu}$ but a less constraining situation for $C_{10\mu}$. This can be improved by considering additional $q^2$ bins and thus over-constraining the system.

It is interesting to consider the $C_{9\mu}$-only and $C_{9\mu}=-C_{10\mu}$ scenarios to show the possible precision of such an over-constrained fit. For the $C_{9\mu}$-only scenario, we use benchmark point 1 of~\eqref{eq:staranddiamond}, which gives
\begin{align}\label{eq:input_scenario1}
\mathcal{A}_{\rm CP}^{\rm dir}[8,9] &= 0.15 \pm 0.02\ , & \mathcal{A}_{\rm CP}^{\rm mix}[1.1,6] &= 0.66 \pm0.04 \ ,  \\
\mathcal{B}[1.1,6.0] &= (1.16\pm 0.02)\times 10^{-7} \ , &  \mathcal{B}[15,22] &= (0.806\pm 0.016)\times 10^{-7} \nonumber \ ,
\end{align}
where the uncertainties again indicate a possible future scenario. Using the inputs in \eqref{eq:input_scenario1} for $C_{9\mu}$ only and \eqref{eq:input} for $C_{9\mu}=-C_{10\mu}$, we find the $68
\%$ and $90\%$ C.L. regions in Fig.~\ref{fig:fitresultscons}.

\begin{figure}[t]
    \centering
    \includegraphics[width=0.6\textwidth]{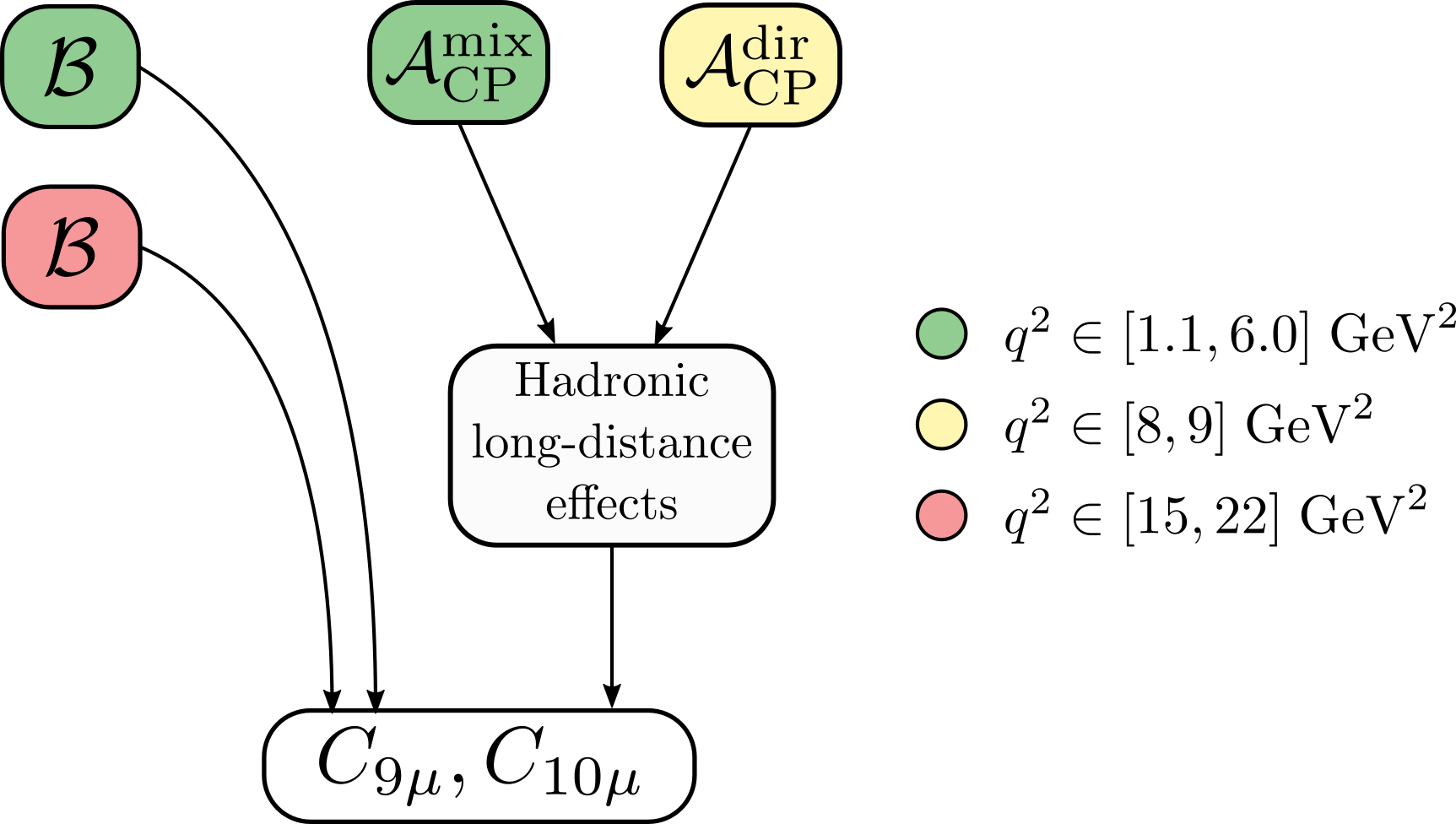}
    \caption{Illustration of the strategy to determine complex Wilson coefficients $C_{9\mu}$ and $C_{10\mu}$.}
    \label{fig:flowchart}
\end{figure}

Finally, we illustrate our strategy in Fig. \ref{fig:flowchart}. By combining the direct and mixing-induced CP asymmetries with the branching ratios in specific bins, we can optimally exploit the complementarity of these observables. In this way, both the hadronic long-distance effects as well as the short-distance Wilson coefficients can be determined from the data.

\section{Testing lepton flavour universality}
\label{ch:RK}
\subsection{Setting the stage}
Previously, we considered new sources of CP violation specifically for the muonic channel. In the SM, the only difference between the muon and electron channels is caused by tiny phase space effects, while $C_9$ and $C_{10}$ are strictly lepton-flavour universal. It is possible that NP enters in a lepton-flavour universal way as in the SM. On the other hand, NP may also affect the different generation distinctly (see e.g. \cite{Buttazzo:2017ixm,Bordone:2019uzc,Bordone:2017bld,Bordone:2018nbg, Cornella:2021sby}). When testing for lepton-flavour universality, special care is needed because new sources of CP violation may play a role. To demonstrate this, we define the integrated rates as
\begin{equation}
    \bar{\Gamma}_\ell[q^2_{\rm min}, q^2_{\rm max}] \equiv  \int_{q^2_{\rm min}}^{q^2_{\rm max}} dq^2 \frac{ d \Gamma(B^- \to K^- \mu^+\mu^-)}{dq^2}   
\end{equation}
and equivalently for $\Gamma_\ell$ for the $B^+\to K^+\mu^+\mu^-$. For simplicity, we omit the explicit $q^2$-bin in the following. Deviations from lepton flavour universality between muons and electrons are then probed through the ratios

\begin{equation}
 R_K \equiv \frac{\Gamma_\mu}{\Gamma_e} \ , \quad\quad\quad  \bar{R}_K \equiv \frac{\bar{\Gamma}_\mu}{\bar{\Gamma}_e} \ ,
    \label{eq:RK}
\end{equation}
for the $B^+$ and its conjugate $B^-$ mode, respectively. 

We note that, usually, the CP-averaged ratio is quoted as $R_K$, although this is often not explicitly written. To make this difference explicit, we define
\begin{equation}
   \langle R_K \rangle  = \frac{ \Gamma_\mu +\bar{\Gamma}_\mu}{\Gamma_e + \bar{\Gamma}_e} \ .
    \label{eq:RKav}
\end{equation}
We note that this averaged quantity is not the same as the sum of $R_K$ and $\bar{R}_K$ as defined in \eqref{eq:RK} but rather
\begin{equation}
    \langle R_K \rangle = \frac{1}{2} \left[R_K + \bar{R}_K + (\bar{R}_K - R_K) \mathcal{A}_{\rm CP,e}^{\rm dir}\right] \,
\end{equation}
where the direct CP asymmetry for the electron mode is defined analogously to the muon mode in \eqref{eq:ACP_dir}. In the SM, $R_K$ and $\bar{R}_K$ are $1$ to an excellent precision, even when including tiny QED effects \cite{Bordone:2016gaq, Isidori:2022bzw}. 

The latest LHCb measurements in the $q^2 \in [1.1,6.0] \: \si{\giga\eV^2}$-bin reads \cite{LHCb:2021trn} 
\begin{equation}\label{eq:rkmeas}
    \langle R_K \rangle[1.1,6.0] = 0.846^{+0.044 }_{-0.041} \ .
\end{equation}
This measurement deviates from the SM prediction with a significance of $3.1 \sigma$ \cite{LHCb:2021trn}. Measurements of this quantity have also been given by the BaBar \cite{BaBar:2012mrf} and Belle \cite{BELLE:2019xld} collaborations. 

Using the LHCb measurement of the muon channel in \eqref{eq:adirmeas}, combined with \eqref{eq:rkmeas} gives \cite{LHCb:2021trn}:
\begin{equation}
    \mathcal{B}(B\to K e^+e^-)[1.1,6.0] = (1.40 \pm 0.10) \times 10^{-7} \ .
\end{equation}
As said, the SM prediction only differs from the the muonic channel in \eqref{eq:SM_BR_lowq2} through tiny phase space effects. Taking \eqref{eq:SM_BR_lowq2}, we find
\begin{equation}
\frac{   \mathcal{B}(B^- \to K^- e^+ e^-)^{\rm SM}[1.1,6.0]_{\rm {{incl/hybrid}}}}{\mathcal{B}(B^\pm \to K^\pm e^+ e^-) [1.1, 6.0]} -1 = 0.31 \pm 0.14 \ , 
\end{equation}
differing by $2.2\sigma$. A somewhat larger difference was found in \cite{Parrott_SM_predictions}. Using the exclusive CKM factors, we find 
\begin{equation}
\frac{   \mathcal{B}(B^- \to K^- e^+ e^-)^{\rm SM}[1.1,6.0]_{\rm excl.}}{\mathcal{B}(B^\pm \to K^\pm e^+ e^-) [1.1, 6.0]} -1 = 0.13 \pm 0.12 \ , 
\end{equation}
which differs by $1.1 \sigma$. We note that even though these SM predictions suffer from uncertainties due to the modeling of the long-distances effects, they seem to indicate consistently that there may be NP present in the muon and/or electron channels.

\subsection{New CP-violating couplings}
If $C_{9\mu}$ is different from $C_{9e}$, they would cause a difference between $R_K$ and $\bar{R}_K$. For completeness, we note that if we have NP in $C_T$, which enters with proportional to the mass, this would also give an effect if $C_{T\mu} = C_{Te}$. However, precisely because they are proportional to the mass, these effects are also suppressed (see previous discussions). Therefore, we do not further discuss such terms. 

We stress that it important to separately measure these two ratios. The amount of CP violation in $R_K$ can be defined as 
\begin{equation}
\mathcal{A}_{\rm CP}^{R_K} \equiv \frac{\bar R_K - R_K}{\bar R_K + R_K} \ .
\label{eq:CPV_in_RK}
\end{equation}
This new observable in \eqref{eq:CPV_in_RK} provides a measure of whether lepton flavour non-universal NP in $B \to K\ell^+\ell^-$ are also CP violating. We can rewrite \eqref{eq:CPV_in_RK} in terms of the direct CP asymmetries of the individual muonic and electronic decay channels:
\begin{equation}
\begin{aligned}
\mathcal{A}_{\rm CP}^{R_K} &=  \left[ \frac{\mathcal{A}_{\rm CP}^{\rm dir,\mu} - \mathcal{A}_{\rm CP}^{\rm dir,e}}{1 - \mathcal{A}_{\rm CP}^{\rm dir,\mu}\ \mathcal{A}_{\rm CP}^{\rm dir,e}} \right] \ ,
\label{eq:CP_separated_RK_direct_CP_asymmetries}
\end{aligned}
\end{equation}
where $\mathcal{A}_{\rm CP}^{\rm dir, \mu}$ and $\mathcal{A}_{\rm CP}^{\rm dir,e}$ denote the direct CP asymmetries of $B \to K\mu^+\mu^-$ and $B \to Ke^+e^-$, respectively defined in \eqref{eq:ACP_dir}. From~\eqref{eq:CP_separated_RK_direct_CP_asymmetries}, we observe that the $\mathcal{A}_{\rm CP}^{R_K}$ will be enhanced if $\mathcal{A}_{\rm CP}^{\rm dir, \mu}$ has an opposite sign to $\mathcal{A}_{\rm CP}^{\rm dir, e}$. In addition, if the muonic and electronic direct CP asymmetries are identical, the observable vanishes. Therefore, any measurement of this observable is a clear sign of CP-violating NP with different magnitude and phase for the electron and muon channels. 

Interestingly, is also possible to access the electronic direct CP asymmetry directly through separate measurements of the ratios for the $B^-$ and $B^+$. Specifically, we find
\begin{equation}
    \mathcal{A}_{\rm CP}^{\rm dir,e} = \frac{2 \langle R_K \rangle - R_K - \bar R_K}{\bar R_K - R_K} = \frac{2 \langle R_K \rangle}{\bar{R}_K - R_K} - \frac{1}{\mathcal{A}_{\rm CP}^{R_K}} \ .
\end{equation}
At the moment, the only limit on the $\mathcal{A}_{\rm CP}^{\rm dir, e}$ comes from the Belle Collaboration\footnote{They use a weighted average over different $q^2$ bins, both below and above the $c \bar c$ resonances.} \cite{Belle:2009zue}:
\begin{equation}
    \mathcal{A}_{\rm CP}^{\rm dir, e} = 0.14 \pm 0.14 \ , 
    \end{equation}
therefore, it would interesting to see if the new observable could be used to obtain a more precise determination.


%
%
%
\section{Conclusions}
\label{ch:conclusions}
The charged and neutral $B\to K \mu^+\mu^-$ decays offer exciting probes of New Physics. In this paper, our main focus has been on imprints of possible new sources of CP violation. We have performed a comprehensive analysis of these channels, using the most recent state-of-the-art lattice QCD calculations for the required non-perturbative form factors. 

For the $q^2$ regions with a large impact of hadronic $c\bar c$ resonances, we have implemented a model by the LHCb collaboration using experimental data. The corresponding fit results in different branches of the hadronic parameters describing the resonances. Interestingly, these effects generate CP-conserving strong phases, which are necessary ingredients for direct CP violation. In the SM, such CP asymmetries are negligibly small. However, new CP-violating phases arising in the short-distance Wilson coefficients may lead to sizeable CP violation, thereby signalling the presence of New Physics. 

In our analysis, we have complemented these direct CP asymmetries with mixing-induced CP violation in the $B^0_d\to K_{S}\mu^+\mu^-$ decay. We have pointed out that the interplay of the CP asymmetries, utilising also information from the differential decay rates, allows us to distinguish between the different hadronic parameter sets describing the resonance effects, as well as between different NP scenarios. In these studies, we have considered the information for the time-dependent differential decay rates integrated over the angle describing the kinematics of the muon pair, thereby simplifying the analysis. Measuring angular distributions would provide even more information. 

We highlight that, in agreement with other studies, we have found that the differential decay rates calculated in the SM are significantly smaller than the experimental values, thereby indicating the presence of NP. We have pointed out that the discrepancies between inclusive and exclusive determinations of $|V_{cb}|$ -- and to a smaller extent of $ |V_{ub}|$ -- have a large impact on the branching ratios. While the exclusive values result in a difference of the charged $B\to K \mu^+\mu^-$ branching ratio in the $q^2$ region within [1.1,6.0]\,$\si{\giga \eV^2}$ at the $2.5\,\sigma$ level, a hybrid scenario pairing the inclusive value of $|V_{cb}|$ with the exclusive value of $|V_{ub}|$ results in a SM branching ratio lying $3.5\,\sigma$  below the measured result. The latter CKM combination gives the most consistent picture of constraints on neutral $B$ mixing with the Standard Model. It will be important to finally resolve the issues in the determination of these CKM matrix elements.

We have presented a new strategy to determine the complex Wilson coefficients $C_{9\mu}$ and $C_{10\mu}$ from measurements of the branching ratios as well as the direct and mixing-induced CP asymmetries in appropriate $q^2$ regions. The method makes use of our finding that different NP sources lead to distinct “fingerprints”, allowing a transparent determination of these coefficients without making specific assumptions, such as having NP only in $C_{9\mu}$ or assuming the relation $C_{9\mu}^{\rm NP}=-C_{10\mu}^{\rm NP}$, as is frequently done in the literature. Since not all required measurements are yet available, we have illustrated this method through specific examples.

In the presence of NP, we may have a violation of lepton flavour universality which is probed through the $R_K$ ratio. Using the different determinations for the CKM factors, we found that the integrated SM branching ratio of the $B^-\to K^- e^+e^-$ channel for $q^2$ in [1.1,6.0]\,$\si{\giga \eV^2}$ is about $2 \sigma$ below the corresponding experimental result for the hybrid/inclusive case and about $1\sigma$ for the exclusive. We have pointed out that new sources of CP violation require special care in studies to distinguish between $B\to K \ell^+\ell^-$ decays and their CP conjugates for the final states with muons and electrons. We have presented a new method to measure direct CP violation in the $B\to K e^+ e^-$ modes using only $R_K$-like ratios for decays and their CP conjugates. 

It will be very interesting to monitor how the data will evolve in the future. The methods to "fingerprint" CP-violating NP in $B\to K \mu^+\mu^-$ decays offer an exciting playground for the future high-precision era of $B$ physics. It will also be exciting the see whether new sources of CP violation can be revealed in these semileptonic rare $B$ decays, and whether they will complement current puzzles in CP violation in non-leptonic $B$ decays. 

\section*{Acknowledgements}
This research has been supported by the Netherlands Organisation for Scientific Research (NWO).

\appendix

\section{\boldmath Coefficients $\rho_i$}
\label{ch:coefficients}
 In Table \ref{tab:BR_prefactors} we list numerical values for the $\rho_i$ coefficients defined in \eqref{eq:CP_asymm_numerator_numerical} and \eqref{eq:CP_asymm_denominator_numerical} integrated over three different $q^2$ bins. For the coefficients that are sensitive to the choice of long-distance branch, we give separate values for the four different $Y_{\pm \pm}$ branches. For the other coefficients, we sum over the branches. 

\begin{table}[]
    \centering
    \begin{tabular}{llll}\toprule
    Coefficient & \multicolumn{3}{c}{Value per $q^2$ bin ($\times 10^2$ except $\Gamma_{
    \rm SM}$)}\\
    \cmidrule(lr){2-4}
    & $[1.1-6.0] $ & $[8-9]$ & $[15-22]$\\\midrule
    $\Gamma_{\rm SM}$ & $(7.34 \pm 0.58) \times 10^{-20}$& $(1.63 \pm 0.19) \times 10^{-20}$& $(5.61 \pm 0.50) \times 10^{-20}$\\
    $\rho_9/\Gamma_{\rm SM}$ & $3.20 \pm 0.04$ & $2.80 \pm 0.27$& $2.67 \pm 0.10$\\
    $\rho_{10}/\Gamma_{\rm SM}$ & $3.20 \pm 0.04$& $2.80 \pm 0.27$& $2.68 \pm 0.10$\\
    $\rho_P/\Gamma_{\rm SM}$ &$1.03 \pm 0.05$& $2.41 \pm 0.28$& $9.81 \pm 0.79$\\
    $\rho_S/\Gamma_{\rm SM}$ &$1.02 \pm 0.05$& $2.40 \pm 0.28$& $9.78 \pm 0.79$\\
    $\rho_T/\Gamma_{\rm SM}$ &$0.702 \pm 0.078$& $1.44 \pm 0.20$& $2.79 \pm 0.33$\\\midrule
    $\rho_{9\rm Re}/\Gamma_{\rm SM}$ &$22.6 \pm 0.15$& $19.3 \pm 1.7$& $22.4 \pm 0.4$\\
    $\rho_{10\rm Re}/\Gamma_{\rm SM}$ & $-27.6 \pm 0.3$& $-24.1 \pm 2.3$& $-23.1 \pm 0.8$\\
    $\rho_{P\rm Re}/\Gamma_{\rm SM}$ & $-2.20 \pm 0.11$& $-2.16 \pm 0.25$& $-4.08 \pm 0.33$\\
    $\rho_{T\rm Re}/\Gamma_{\rm SM}$ & $1.24 \pm 0.08$& $1.06 \pm 0.11$& $1.22 \pm 0.08$\\
    $\rho_{10P\rm Re}/\Gamma_{\rm SM}$ &$0.509 \pm 0.025$& $0.502 \pm 0.058$& $0.948 \pm 0.077$\\
    $\rho_{9T\rm Re}/\Gamma_{\rm SM}$ & $0.351 \pm 0.024$& $0.306 \pm 0.035$& $0.290 \pm 0.022$\\
    $\rho_{10P\rm Im}/\Gamma_{\rm SM}$ &$-0.509 \pm 0.025$ & $-0.501 \pm 0.059$& $-0.949 \pm 0.077$\\
    $\rho_{9T\rm Im}/\Gamma_{\rm SM}$ & $-0.351 \pm 0.024$& $-0.306 \pm 0.036$& $-0.289 \pm 0.022$\\\midrule
    $\rho_{9\rm Im}/\Gamma_{\rm SM}$($Y_{--}$) & $-3.59 \pm 0.78$& $-13.2 \pm 3.1$& $-2.09 \pm 1.09$\\
    $\rho_{9\rm Im}/\Gamma_{\rm SM}$($Y_{-+}$) &$-2.02 \pm 0.78$& $-10.7 \pm 3.3$& $-2.59 \pm 0.83$\\
    $\rho_{9\rm Im}/\Gamma_{\rm SM}$($Y_{+-}$) &$ 1.25 \pm 0.77$& $9.66 \pm 3.23$& $-1.65 \pm 0.86$\\
    $\rho_{9\rm Im}/\Gamma_{\rm SM}$($Y_{++}$) & $2.80 \pm 0.80$& $12.3 \pm 3.3$& $-5.85 \pm 0.76$\\
    $\rho_{T\rm Im}/\Gamma_{\rm SM}$($Y_{--}$) & $0.197 \pm 0.045$& $0.723 \pm 0.176$& $-0.112 \pm 0.059$\\
    $\rho_{T\rm Im}/\Gamma_{\rm SM}$($Y_{-+}$) & $0.111 \pm 0.045$& $0.586 \pm 0.187$& $0.141 \pm 0.047$\\
    $\rho_{T\rm Im}/\Gamma_{\rm SM}$($Y_{+-}$) & $-0.0673 \pm 0.0436$ & $-0.524 \pm 0.187$& $0.0889 \pm 0.0477$\\
    $\rho_{T\rm Im}/\Gamma_{\rm SM}$($Y_{++}$) &$-0.155 \pm 0.045$& $-0.680 \pm 0.182$& $0.319 \pm 0.047$\\
    \bottomrule
    \end{tabular}
    \caption{Numerical values of the coefficients in Appendix \ref{ch:coefficients}. $q^2$ ranges are given in units of $\si{\giga \eV^2}$. All results are specific to $B^\pm \to K^\pm \mu^+\mu^-$.} 
    \label{tab:BR_prefactors}
\end{table}

\section{Analytical solution}
\label{ch:analytical_solution}
In this section, we derive the analytic expressions for $C_9$ and its phase $\phi_9$ in terms of the mixing-induced CP asymmetry and the direct CP asymmetry assuming either NP only in $C_{9\mu}$ or in $C_{9\mu} = -C_{10\mu}$.  To do this, we use \eqref{eq:Adir_numerator} and \eqref{eq:s0}, describing the numerators of the direct and mixing-induced CP asymmetries. To solve the system, we write:
\begin{equation}
\begin{aligned}
    s_0(q^2) &= s_0^{\rm SM}(q^2) + a(q^2) \abs{C_9^{\rm NP}} \cos \phi_9 + b(q^2) \abs{C_9^{\rm NP}} \sin \phi_9 + c(q^2) \abs{C_9^{\rm NP}}^2 \bigg(\cos^2 \phi_9 - \sin^2 \phi_9 \bigg)\\
    &+ d(q^2) \abs{C_9^{\rm NP}}^2 \sin\phi_9 \cos\phi_9 + e(q^2) \abs{C_{10}^{\rm NP}} \left(\abs{C_{10}^{\rm SM}} \cos \phi_{10} + \frac{1}{2} \abs{C_{10}^{\rm NP}} \bigg(\cos^2 \phi_{10} - \sin^2 \phi_{10} \bigg)  \right) \\
    &+  f(q^2) \abs{C_{10}^{\rm NP}} \sin \phi_{10} \left( \abs{C_{10}^{\rm SM}} + \abs{C_{10}^{\rm NP}} \cos\phi_{10}\right) \ ,
\end{aligned}
\end{equation}

with 
\begin{equation}
\begin{aligned}
    s_0^{\rm SM} =& \mathcal{C}_n \bigg\{  \frac{\lambda_B}{6 q^2} (1 + 2 \hat m_\ell^2) \bigg( \bigg(\frac{2 m_b}{m_B + m_K} C_7 f_T(q^2)\bigg)^2\\
    &+ \frac{4 m_b}{m_B + m_K} C_7 f_T(q^2) f_+(q^2) \bigg(C_9^{\rm SM} + \Re Y(q^2) \bigg)\\
    &+ f_+^2(q^2) \bigg[(C_9^{\rm SM})^2 + 2 C_9^{\rm SM} \Re Y(q^2) + \abs{Y(q^2)}^2\bigg] \bigg)\\
    &+ \bigg( \frac{\lambda_B}{6 q^2} \beta_\ell^2 f_+^2(q^2) + (m_B^2 - m_K^2)^2 \frac{m_\ell^2}{q^4} f_0^2(q^2) \bigg) (C_{10}^{\rm SM})^2\bigg\} \ ,
\end{aligned}
\end{equation}
\begin{align}
    a(q^2) & =    \mathcal{C}_n \frac{\lambda_B}{3 q^2} (1 + 2 \hat m_\ell^2) \bigg( \frac{2 m_b}{m_B + m_K} C_7 f_T(q^2) f_+(q^2)+  f_+^2(q^2) (C_9^{\rm SM} + \Re Y(q^2)) \bigg) \ , \nonumber \\
 b(q^2)  &  = a(\mathcal{C}_n\to \hat{\mathcal{C}}_n ) \ ,\nonumber \\
 c(q^2)   & = \mathcal{C}_n  \frac{\lambda_B}{6 q^2} (1 + 2 \hat m_\ell^2) f_+^2(q^2) \ ,\nonumber \\
    d(q^2) & =   c(\mathcal{C}_n\to   \hat{ \mathcal{C}}_n ) \ ,\nonumber \\
  e(q^2)  &  = \mathcal{C}_n \bigg( \frac{\lambda_B}{3 q^2} \beta_\ell^2 f_+^2(q^2) + 2(m_B^2 - m_K^2)^2 \frac{m_\ell^2}{q^4} f_0^2(q^2) \bigg)  \ ,\nonumber \\
 f(q^2)  &  = e(\mathcal{C}_n\to \hat{\mathcal{C}}_n) \ ,
\end{align}
where we also took the small CKM phase in the normalization into account via \begin{equation}
    \mathcal{C}_n \equiv \bigg[-4 \cos \phi_d \Re \mathcal{N} \Im \mathcal{N} -2 \sin \phi_d ((\Re \mathcal{N})^2 - (\Im \mathcal{N})^2)\bigg] \ ,
\end{equation}
and
\begin{equation}
   \hat{ \mathcal{C}}_n \equiv  \bigg[+4 \sin \phi_d \Re \mathcal{N} \Im \mathcal{N} -2 \cos \phi_d ((\Re \mathcal{N})^2 - (\Im \mathcal{N})^2) \bigg]  \ .
\end{equation}

Taking now the differential rate for the $B\to K_S$\footnote{The charged rate can also be used by taking into account the normalization factor.}, we find that 
\begin{equation}
    \abs{C_9^{\rm NP}} = \frac{(\bar \Gamma - \Gamma)}{\rho_{9 \rm Im} \sin \phi_9^{\rm NP}} =  \frac{2\mathcal{B}(B_d^0\to K_S\ell^+\ell^-) \mathcal{A}_{\rm CP}^{\rm dir} \tau_B}{\rho_{9 \rm Im} \sin \phi_9^{\rm NP}} \ .
\end{equation}
Finally, we can solve the system of equations. For $C_{10}^{\rm NP} =0$, we then find
\begin{align}\label{eq:phisol}
    &\cot \phi_9^{\rm NP} = - \frac{1}{2c}\bigg(a \frac{\rho_{9 \rm Im}}{(\bar \Gamma - \Gamma)} + d\bigg) \nonumber \\
     &\pm \sqrt{\frac{1}{(2c)^2}\bigg(a \frac{\rho_{9 \rm Im}}{(\bar \Gamma - \Gamma)} + d\bigg)^2 - \bigg(\frac{(s_0^{\rm SM} - s_0)}{c} \frac{\rho_{9 \rm Im}^2}{(\bar \Gamma - \Gamma)^2} + \frac{b}{c} \frac{\rho_{9 \rm Im}}{(\bar \Gamma - \Gamma)} - 1 \bigg)} \ .
\end{align}
For the scenario in which $C_9= - C_{10}$, the solution is obtained from \eqref{eq:phisol} by replacing: 
\begin{equation}
d\to d -f \ , \quad a \to a + C_{10}^{\rm SM} e \ , \quad b \to b - C_{10}^{\rm SM} f   \ ,  \quad c \to c + \frac{e}{2} \ .
\end{equation}

\clearpage

\bibliographystyle{JHEP} 
\bibliography{refs.bib}
\end{document}